\documentclass[a4paper]{article}
\usepackage{setspace}
\usepackage{graphicx} 
\usepackage{amsmath}
\usepackage{amssymb}
\usepackage[backend=biber,style=numeric,sorting=none]{biblatex}
\usepackage{csquotes}
\usepackage{caption}
\usepackage{wrapfig}
\usepackage{subcaption}
\usepackage{color}
\usepackage{pstricks}
\usepackage{multirow}
\usepackage{pst-barcode}
\newcommand{\argmin}{\mathop{\mathrm{arg\,min}}}

\newcommand{\R}{\mathbb{R}}

\title{Mathematical framework for perception-driven parameter choice in image denoising}

\usepackage{authblk} 
\author[1]{Saara Isoranta}
\author[2]{Emilia L.K. Bl\r{a}sten}
\author[ \hspace{-1ex}]{Lílian Ferreira de Freitas}
\author[3]{Jukka H\"akkinen}
\author[1]{Markus Juvonen}
\author[1]{Samuli Siltanen}

\affil[1]{Department of Mathematics and Statistics, University of Helsinki, Helsinki, Finland}
\affil[2]{Department of Computational Engineering, LUT University, Lahti, Finland}
\affil[3]{Faculty of Technology and Seafaring, Maritime Technology, Novia University of Applied Sciences, Turku, Finland}

\begin{document}

\maketitle




\section{Introduction}





\begin{figure}
\begin{picture}(300,140)
    \put(3,87){\includegraphics[width=12cm]{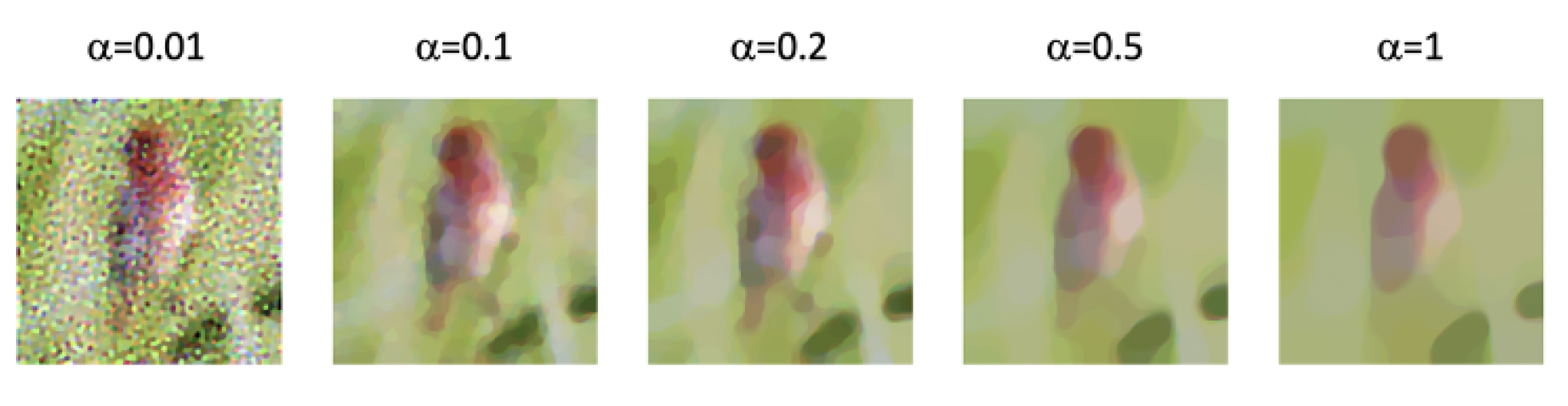}}
    \put(0,-13){\includegraphics[width=12cm]{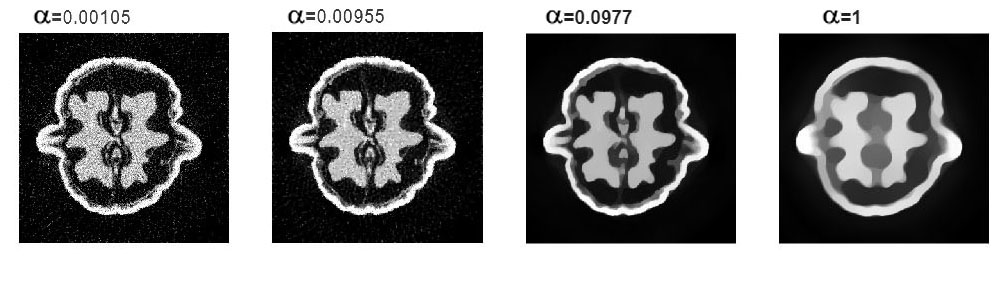}}
\end{picture}
\caption{Two inverse problems solved with total variation (TV) regularization by substituting a discrete version of $\mathcal{R}(f)=\|\nabla f\|_1$ to (\ref{eq:inverse_problem}). Note the smoothing effect of increasing regularization parameter $\alpha$. Top row: deblurring a photograph of a bird. Bottom row: X-ray tomography of a walnut.  }
    \label{Fig:lintupic}
\end{figure}

A classical example of applied inverse problems is to recover a clean image $f$ from a noisy observation $\mathcal{A}f$. The variational regularization solution is calculated as the minimizer
\begin{equation}\label{eq:inverse_problem}
    \argmin_f \lbrace \|\mathcal{A}f-m\|^2 + 
    \alpha \mathcal{R}(f) \rbrace
\end{equation}
of a penalty consisting of two terms. The first term $\|\mathcal{A}f-m\|^2$ measures how accurately $f$ reproduces the measured data $m$; the forward operator $\mathcal{A}$ is a model for the measurement. The regularizer $\mathcal{R}$ offers a way to incorporate {\it a priori} information into the solution; the second term $\mathcal{R}(f)$ should give a large number for any unexpected image $f$. The need for regularization comes from possible modeling errors, measurement noise, or general ill-posedness \cite{kaipio05_statis}. The regularizer $\mathcal{R}$ can be an analytic formula or a learned algorithm (see \cite{burger, bungert, scherzer, vanleeuwen} for examples). The parameter $\alpha>0$ tunes a balance between the two terms. See Figure \ref{Fig:lintupic} for an illustration of parameter choice in the cases of image deblurring and X-ray tomography, both covered by equation (\ref{eq:inverse_problem}) with different forward operators $\mathcal{A}$.

In mathematical literature, a variety of solution methods have been proposed for optimal choice of  $\alpha$, such as the L--curve method (early references include \cite{miller, lawson}), the more novel S--curve method (section $4$ in \cite{hamalainen}, \cite{niinimaki}), and Morozov's discrepancy principle (\cite{morozov}, applied in \cite{pereverzyev} for instance). Statistical methods such as Bayesian approaches \cite{kolehmainen}, or the $\chi^2$ test \cite{pearson, mchugh, geeksforgeeks} can also be applied, as well as experimentation or prior knowledge. One can assume that machine learning and artificial intelligence will be used for parameter selection more and more often in the future. It is evident from the wide variety of solution methods in use that the ideal choice of parameter is generally not objective, easy, or obvious.

Our premise is that optimal parameter choice depends on the application area. More specifically, we look for a way to choose the $\alpha$ that best serves the end-user in their imaging task. Therefore, instead of a general mathematical formula, we propose creating an empirical approach based on human visual perception.

\begin{figure}

\begin{picture}(320,170)(-30,0)
\put(240,120){\includegraphics[width=3cm]{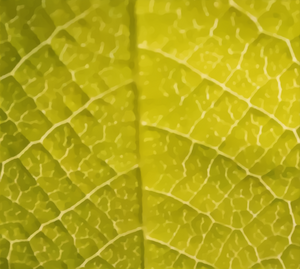}}
\put(150,120){\includegraphics[width=3cm]{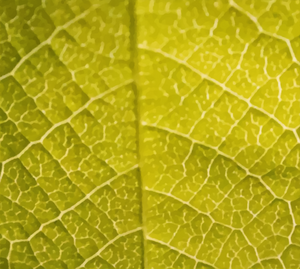}}
\put(60,120){\includegraphics[width=3cm]{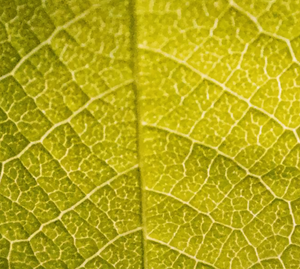}}
\put(-30,120){\includegraphics[width=3cm]{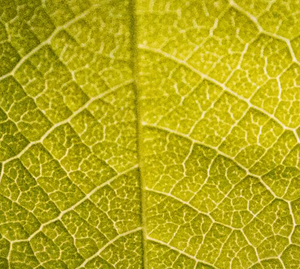}}
\put(240,30){\includegraphics[width=3cm]{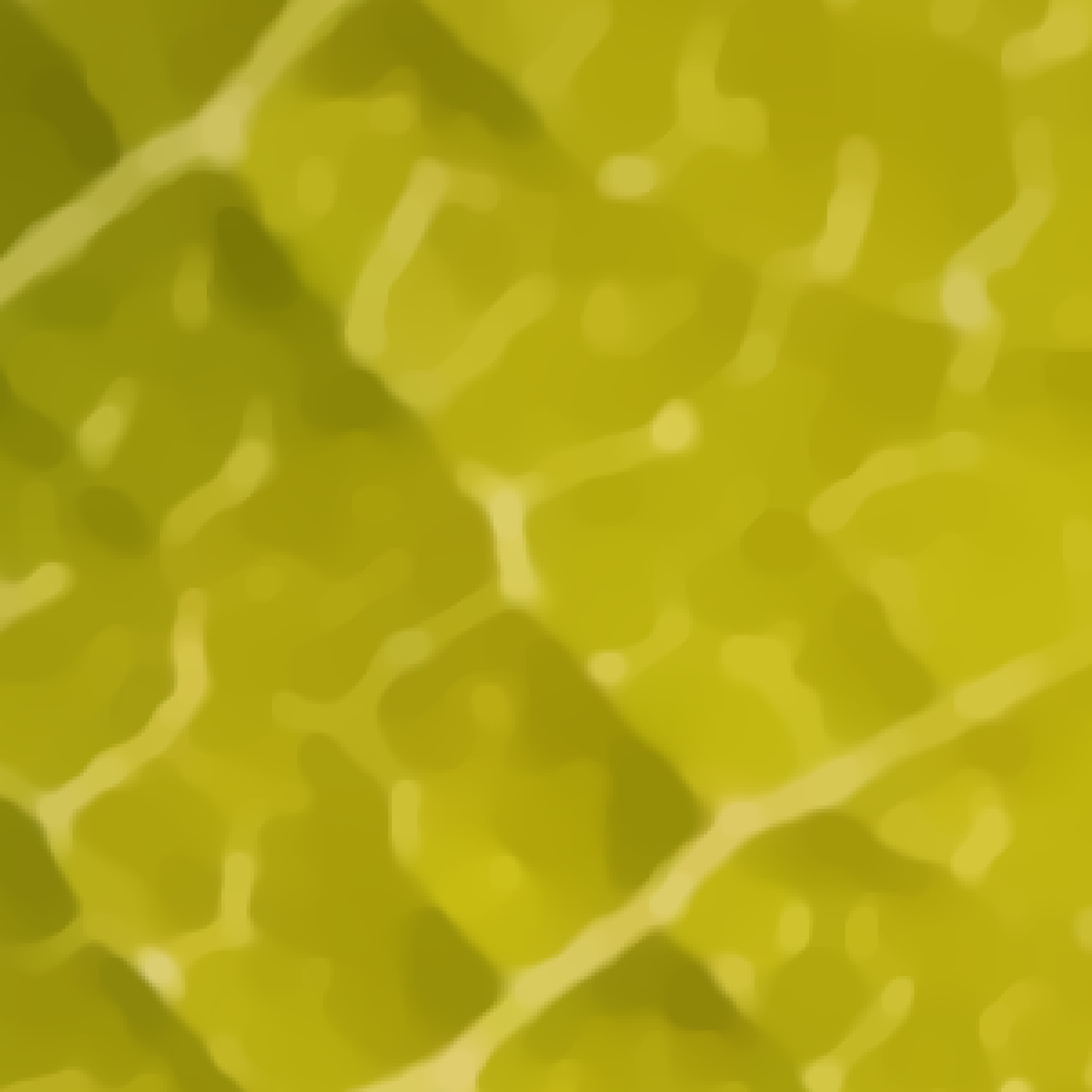}}
\put(240,20){\small $\alpha = 2000$}
\put(150,30){\includegraphics[width=3cm]{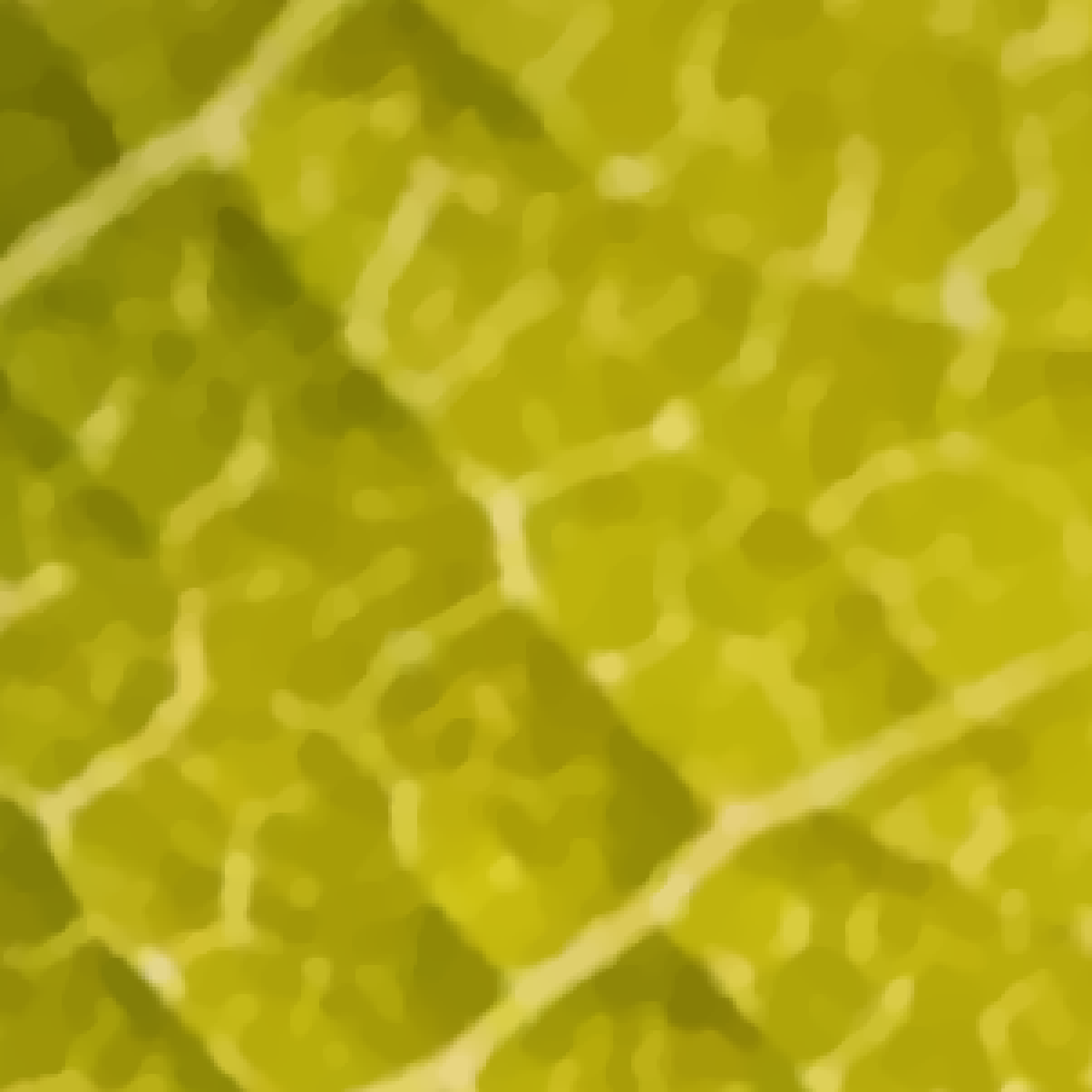}}
\put(150,20){\small $\alpha = 160$}
\put(60,30){\includegraphics[width=3cm]{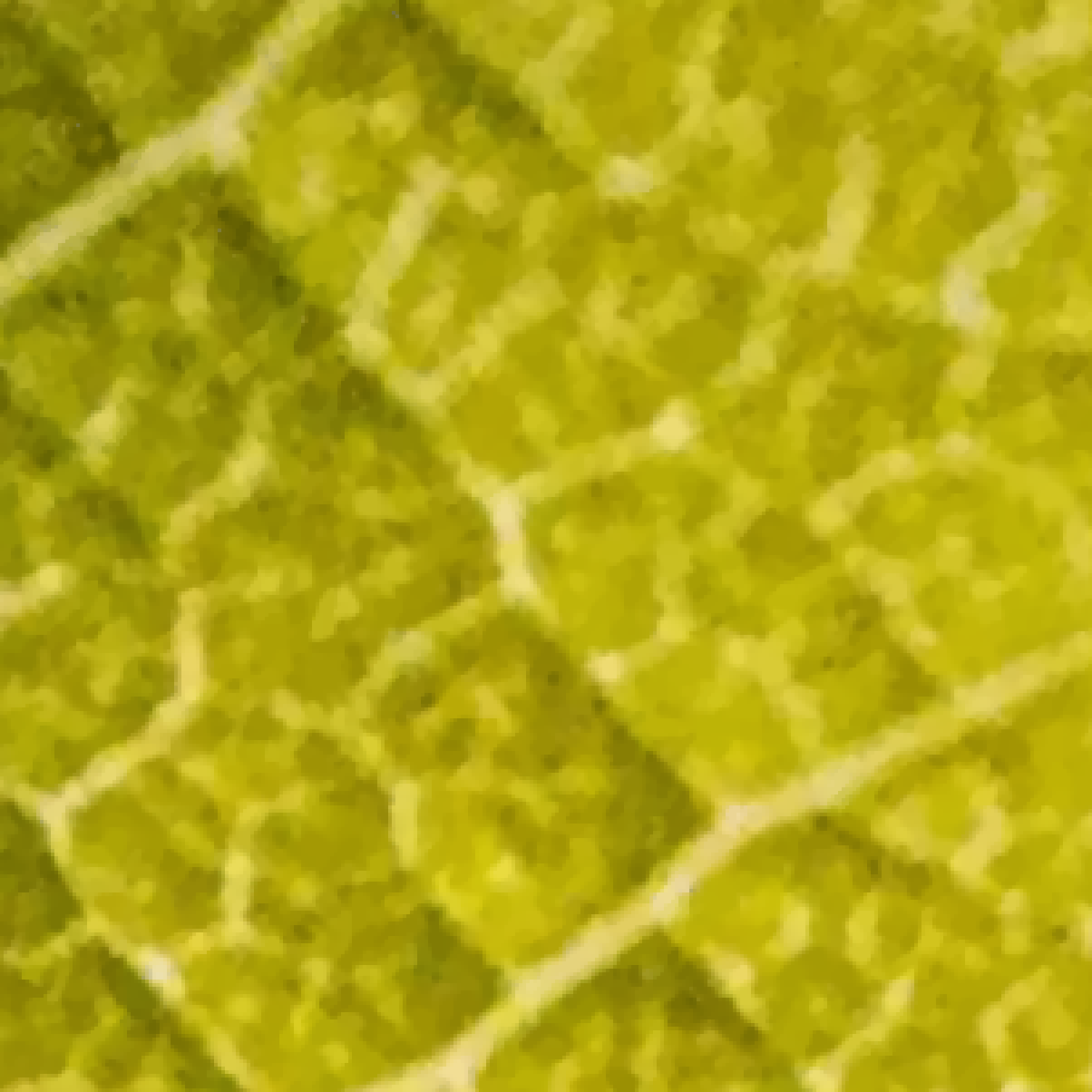}}
\put(60,20){\small $\alpha = 32$}
\put(-30,30){\includegraphics[width=3cm]{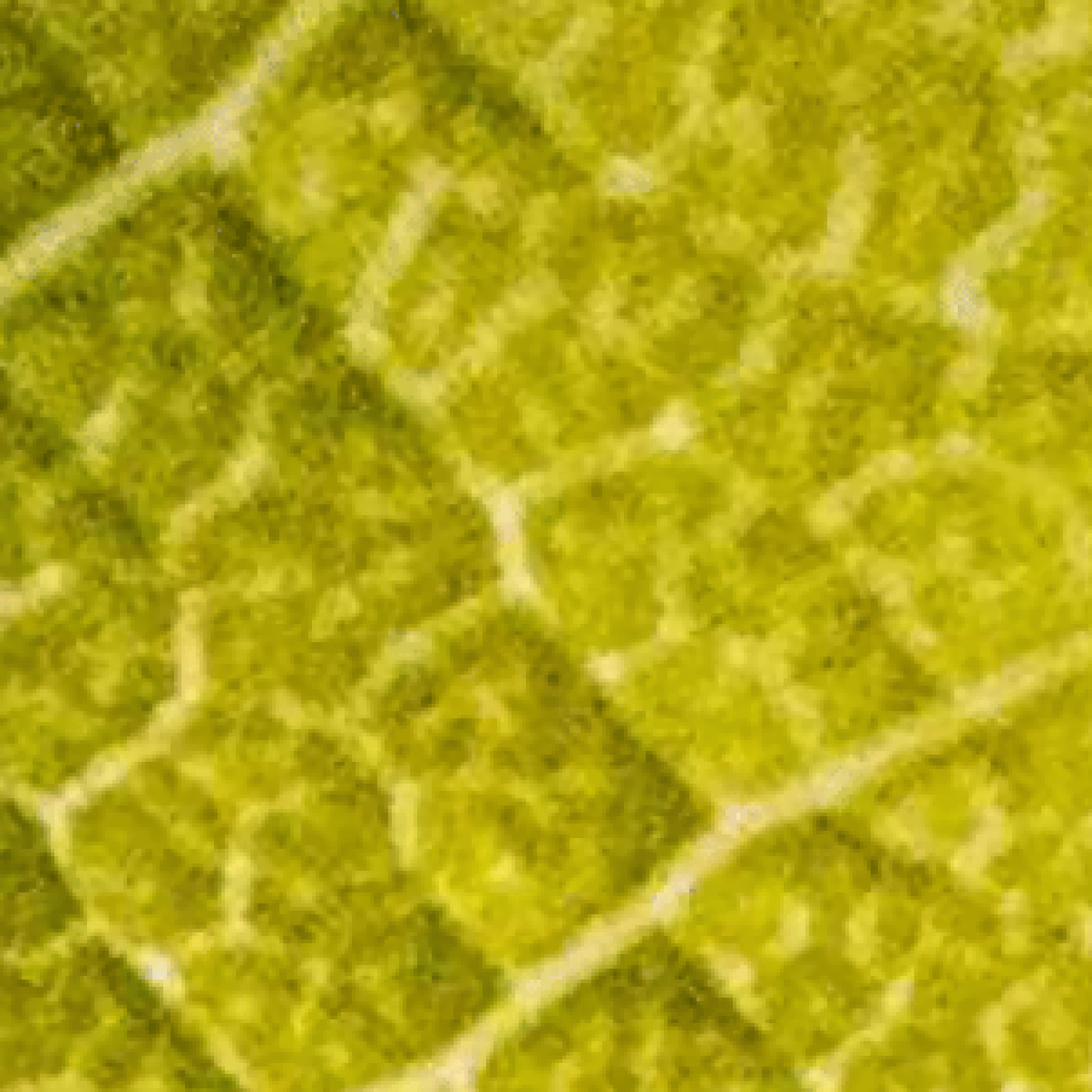}}
\put(-30,20){\small $\alpha = 8$}
\end{picture}

\caption{\label{fig:im_grid}The effect different parameter values have on the same measured data. The differences in strength of denoising are clearly visible in the magnified images in the bottom row. This is an example of denoising using equation (\ref{eq:chambolle-pock}), where the forward operator is identity.}
\end{figure}
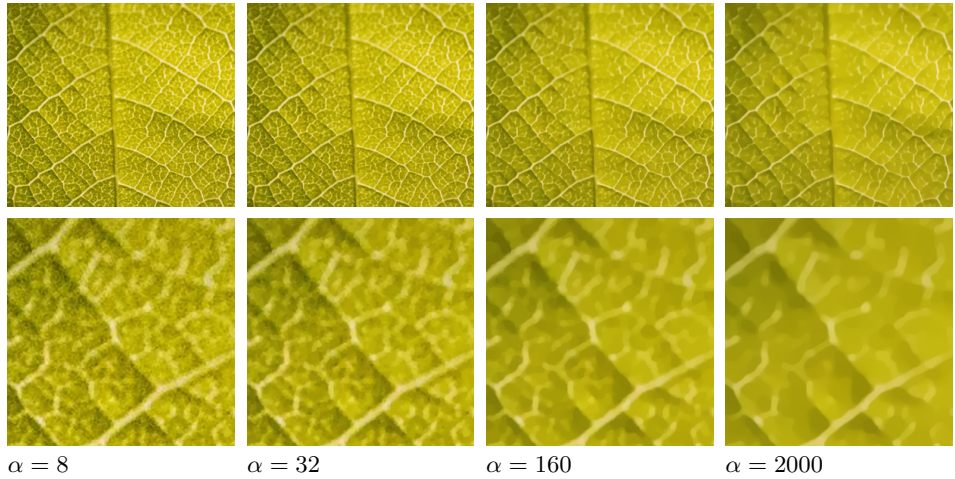

This idea is quite new, so we start by studying a simplified situation. Instead of the more general problem (\ref{eq:inverse_problem}), we let $\mathcal{A}$ be the identity operator and focus on simply denoising an image:
\begin{equation}\label{eq:chambolle-pock}
    f_{\alpha} = \argmin_f \lbrace \|f-m\|^2 + \alpha \| \nabla f\| \rbrace. 
\end{equation}
The choice $\mathcal{R}(f)=\| \nabla f\|$ leads to total variational denoising \cite{rudin, chambolle-pock, bredies}.
Some examples of minimizers of (\ref{eq:chambolle-pock}) with different values of $\alpha$ can be seen in Figure \ref{fig:im_grid}.

So, how to design an experiment to determine which value of $\alpha$ human participants would find to be the optimal one in (\ref{eq:chambolle-pock})? A mathematician's approach could be to display a denoised image on a screen, equipped with a slider to adjust the parameter value, and to ask participants to find the ideal image by operating the slider. This avenue of research is challenging in terms of both analysis and construction of the sample set. Furthermore, mathematical research practices are a poor fit for studies that rely on vision and opinion. In order to ensure that our results truly reflect how people \emph{in general} experience these images, we should approach this issue from the perspective of the science of human perception.

\begin{figure}[ht]
    \centering
    \includegraphics[width=0.9\textwidth]{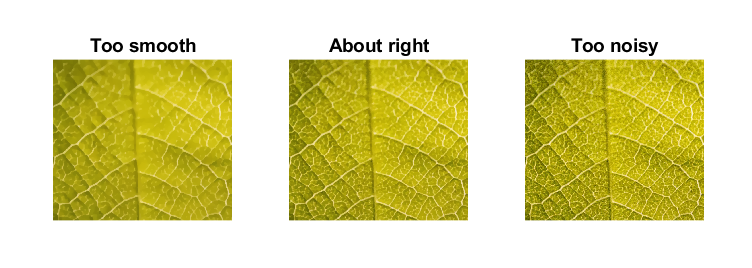}
    \caption{One of the anchor images used in the last test block. This image was only used in in--person testing and accompanied with a verbal instruction.}
    \label{fig:anchor}
\end{figure}

The tried-and-true method for psychological image perception experiments is pairwise comparison \cite{thurstone, engeldrum, bianchi, bockenholt}. Two images are shown to the  participant side by side, and they are asked to choose one of them as an answer to a carefully chosen question. For example, \emph{Which image is noisier than the other?} Then, after collecting a large set of answers, one could analyze them statistically according to \cite{engeldrum}. Of course, the experiment needs to be designed carefully. Researchers must construct a test environment and prepare instructions for the participants. Figure \ref{fig:anchor} displays an anchor image, which is an example of the latter. Most importantly, the appropriate image to use as stimulus must be determined, and we need to conclude which are the possible options for the ideal $\alpha$. In other words, we must construct a parameter grid with which we should compute the sample set of denoised images $f_{\alpha}$ (\ref{eq:chambolle-pock}) before we can move on to searching for the ideal solution.

We should have a grid 
\begin{equation}\label{eq:grid_from_intro}
   \mathcal{G} =  \lbrace \alpha_1, \alpha_2, \cdots , \alpha_{K-1}, \alpha_K \rbrace \subset\R
\end{equation}
of $K$ positive parameters ordered in such a way that for any $k < \ell$ we have $\alpha_k < \alpha_\ell$. Furthermore, we need to calculate denoised images $f_{\alpha_k}$ as minimizers of (\ref{eq:chambolle-pock}) for all $k=1,2,3,\dots K$, which will be used in pairwise comparisons. There are several requirements for the grid $\mathcal{G}$:
\begin{itemize}
\item[(i)] $\alpha_1>0$ should be small enough to make sure that any smaller parameters than $\alpha_1$ surely result in overly noisy images that are unusable in the application.
\item[(ii)] $\alpha_K>0$ should be large enough to make sure that any larger parameters than $\alpha_K$ surely result in over-smoothed images that are unusable in the application. 
\item[(iii)] Uniformly spaced grid may not be the best choice; logarithmic or some other type of non--uniform scaling might be better. Some sources (such as \cite{thurstone}) suggest a normally distributed scale.
\item[(iv)] The grid should not have too many elements because one participant can only do so many comparisons before getting tired. Therefore, the number of grid points should be as small as possible while consecutive elements in the grid should yield reasonably different denoised images.  
\item[(v)] Denoised images $f_{\alpha_k}$ and $f_{\alpha_{k+1}}$ constructed with consecutive grid elements should not be too different from each other. Otherwise there is a risk that the optimal parameter is between   $\alpha_k$ and $\alpha_{k+1}$ and neglected in the process. 
\end{itemize}

The construction of a good grid $\mathcal{G}$ is a challenging problem in itself. Therefore, this paper is devoted to the design of an empirical procedure for creating the grid for further studies. We illustrate how paired comparison experiments can be used to create a grid $\mathcal{G}$ of the form (\ref{eq:grid_from_intro}) satisfying (i)--(v). The focus will be on requirements (iv) and (v), which can be viewed as the defining qualities of the grid $\mathcal{G}$. Requirements (i) and (ii) are rather trivial since they are easily met, and (iii) is a possible consequence of (iv) and (v) rather than an instruction for constructing the grid; it is, however, good to remember that we shouldn't limit our research to uniformly spaced grids.

Part (v) implies that the grid is discretized by the similarity of consecutive denoised images. Presuming that similarity can be measured with both mathematical and psychological methods, we hypothesize that a link can be built between the two. For this, we apply a tool called psychometric scaling \cite{engeldrum}. A grid $\mathcal{G}_0$ based on computed similarity is constructed first, and a comparison test is conducted using the corresponding denoised images as samples. The survey question in this case is if the two images shown are similar or not. Following ``the method of constant stimuli'' \cite{engeldrum}, a numerical threshold for perceived similarity is determined. As a result, we have a constant HaarPSI \cite{reisenhofer} value by which the grid can then be re-discretized so that consecutive images are just noticeably different. This way the grid is not too fine nor too coarse.

Chapter \ref{chapter:psychology} presents the psychological theory and research practices necessary for this study. Chapter \ref{chapter:method} describes the execution and results of the experiment in detail; in chapter \ref{chapter:TV minimization}, the method of denoising images using total variation is introduced, and the important role of the parameter $\alpha$ is explained. The algorithms used in this project are introduced. Chapters \ref{chapter:stimuli} and \ref{chapter: procedure} explain the photographic material and the execution of the experiment. Results are presented in chapter \ref{chapter:results}.

This paper also aims to research and establish methods of collecting observer--based data in a mathematical setting. Comparison tests are a standard tool in perception psychology, and over the course of this study their execution is studied and developed. The experiment base is, as a result, ready for use in further experiments on perception--based imaging.

\section{Psychometric scaling and a just noticeable difference} \label{chapter:psychology}

The concept of similarity is recognized in mathematics primarily as a computational property, while some principles of optics, psychology, engineering, art and personal experience imply that visually perceived similarity is something different. In imaging, it is often measured in pixel--to--pixel correspondence between a solution and what would be the ground truth image (see \cite{reisenhofer, wang} for instance). However, there may be a distinct difference between the highest quality solution by mathematical definition, and the one that is the easiest to interpret visually. We can also assume that we do not experience the world around us without noise. Humans' visual and perceptive abilities vary a lot, and the world in itself contains visual noise in the form of air particles, radiation and other obstacles. A completely noiseless image is likely not the most natural one. Additionally, environmental noise is of different type than the measurement noise caused by photography and the digital process an image goes through. An image denoised of measurement noise may not appear natural.

The field of perception psychology studies the ways in which we process and utilize visual data. Psychometric scaling is an application of perception psychology, that combines the methods of psychological research and statistical analysis. The core principle of psychometric scaling is that humans can perform as meters (chapter $1$, \cite{engeldrum}). As one can deduce from the term, ``psychometric scaling'' stands for the use of psychological or subjective phenomena as a method of measurement. The idea, and the methods used in this paper, originate from the $1927$ article ``A Law of Comparative Judgment'' \cite{thurstone} by L. L. Thurstone. The book ``Psychometric Scaling: A Toolkit for Imaging Systems Development'' \cite{engeldrum} ($2000$) by Peter Engeldrum bases a lot of its contents on Thurstone's article as well. This more applied text was widely used for reference on this paper. 

In his book Engeldrum introduces methods of combining statistical analysis and psychological research as techniques to measure some visual quality in a sample. According to Engeldrum, these qualities - which are referred to as ``-nesses'' - can be categorized and analyzed separately from each other. In photographic imaging and other forms of visual data, the different -nesses can include contrast, blurriness, intensity of color, brightness, focus, and many other visually detectable qualities. The -ness of choice for this study is noisiness.

The goal in this psychometric scaling study is to determine a \emph{just noticeable difference} (JND) in image quality or noisiness. JND, or Weber's law, has been recognized as a concept in psychophysical and perceptual research since the $19$th century \cite{fechner}, but it has not been established in a mathematical setting. The quantity, which is discussed at length in Engeldrum's book as well, can be described as the smallest difference in an attributed value between two samples that is distinguishable to the naked eye. In this paper, we treat JND as a threshold value which can be used to discretize the parameter grid \ref{eq:grid_from_intro}.

Visual image evaluation is the only possible method of data collection in perception-–based research, because so much of the process of perception is still unknown. The complexity of any psychological phenomenon makes any attempt at simulation an unfeasible and unreliable option. The only reliable source at hand for how visual data is perceived are the humans who can report their experience - hence the idea of humans performing as meters.


\section{Materials and methods} \label{chapter:method}

This chapter describes the method in which the experiment was executed. The project was implemented on the scale of a pilot study, and in a full--scale study more rigorous guidelines will be followed. As developing testing methods for mathematical purposed was an important goal of this project, a degree of flexibility and experimentation was accepted in the data collection phase. In order to keep the test site simple and easily modifiable no background information was collected of the participants.

\subsection{Total variation minimization} \label{chapter:TV minimization}


The classical idea behind denoising is to algorithmically ``counteract'' the measurement noise detected in the image.
Total variation (TV) regularization in image processing is a denoising or deblurring method that has an edge--preserving and smoothing quality. The method was launched by Leonid Rudin, Stanley Osher and Emad Fatemi in their $1992$ article, ``Nonlinear total variation based noise removal algorithms'' \cite{rudin}. As its name implies, the principle of TV regularization is to minimize the total variation of the oscillations of a noisy image signal \cite{mallet}.

In this study, we use the Chambolle--Pock algorithm \cite{chambolle-pock}, which is based on a minimizing equation (\ref{eq:chambolle-pock}). It yields good results in total variation, and relies less on having a good a priori approximation than a non-iterative method. The regularization parameter, $\alpha$, is the only adjustable variable in the algorithm. By tuning its value, our goal is to find a suitable solution $f$ such that
\begin{enumerate}
    \item its values are close to the measured data,
    \item its gradient is of reasonable size, and
    \item observers find it a nice compromise.
\end{enumerate}

The parameter $\alpha$ is the focus in our project. As we can see from equation~(\ref{eq:chambolle-pock}), the coefficient $\alpha$ determines the ratio of data fitting to regularization.
Ideally, the best value for the parameter would be such that the resulting image is denoised but not blurry, with clearly visible edges and a sufficient level of detail in texture. When $\alpha$ is large, the functional reaches minimal values when the gradient of $f$ is small. The result is then attenuated and often has an over--smooth, blurry appearance. Then again, when the values of $\alpha$ are small, the minimal values are reached when $f$ is very close to the measured data $m$, meaning that the result faces very little noise removal, often resulting in a grainy or noisy image (see figure \ref{fig:im_grid}).

The images for this study were computed using a Matlab code \cite{c-p, zenodo}, written by Emilia Blåsten and Lílian Ferreira de Freitas, 
of the Chambolle--Pock algorithm \cite{chambolle-pock}, with some ideas from \cite{bredies} also applied in the code. The Matlab function defines the algorithm using a different, but equivalent notation
\[
\arg \min_f\| \nabla f\| + \frac{\lambda}{2} \| f-m\|^2 .
\]
The correspondence between the parameters $\alpha$ and $\lambda$ is $\alpha = 2/\lambda$. The parameters $\lambda$ range from $0.0001$ to $5$, meaning that the values of $\alpha$ range from $0.4$ to $2000$. $200$ iterations were used for each image.

\subsection{Participants}


$15$ graduate students, researchers, and family members of the research team participated in the first block of the experiment in person. $24$ observers participated in the second block. $8$ of these took the test in person and $16$ remotely. The third block had $13$ participants from the same demographic. $20$ observers participated in the fourth block, where participation was expanded to the general public as well. See table \ref{table:data} for a summary.

The age of the participants ranged between $18$ and $60$, with the majority being of approximately $30$ years of age, and all genders were represented.
The participants received no compensation, and were told that the study is about how humans detect differences in images with the naked eye.

\begin{table}
\centering
\begin{tabular}{ |p{1cm}||p{3,5cm}|p{2cm}|p{2cm}|  }
 \hline
 \multicolumn{4}{|c|}{Collected data} \\
 \hline
 Block & Test data & Number of participants & Number of evaluations \\
 \hline
 $1$   & $[f_{\alpha_{16}}, f_{\alpha_{23}}]$    & 15 &   565\\
 $2$ &  $[f_{\alpha_1}, f_{\alpha_8}]$  & 24   & 547\\
 $3$ & $[f_{\alpha_{10}}, f_{\alpha_{14}}], [g_{\beta_{11}}, g_{\beta_{17}}]$ & 13 &  398\\
 $4$ & $[h_{\gamma_{10}},h_{\gamma_{19}}]$ & 20 &  449\\
 \hline
 \end{tabular}
 \caption{A summary of the collected data.}
\label{table:data}
\end{table}

\subsection{Stimuli} \label{chapter:stimuli}

\begin{figure}[ht]
\centering
\begin{subfigure}{0.3\textwidth}
  \centering
  \includegraphics[width=.91\linewidth]{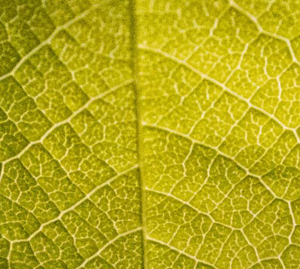}
  \caption{\emph{Image $1$}}
  \label{fig:leaf}
\end{subfigure}
\begin{subfigure}{.3\textwidth}
  \centering
  \includegraphics[width=.86\linewidth]{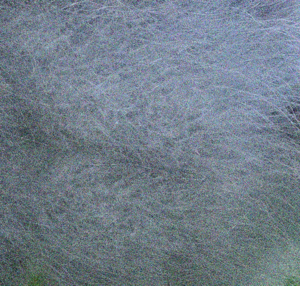}
  \caption{\emph{Image $2$}}
  \label{fig:down}
\end{subfigure}
\begin{subfigure}{.3\textwidth}
  \centering
  \includegraphics[width=.89\linewidth]{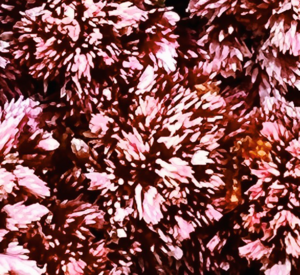}
  \caption{\emph{Image $3$}}
  \label{fig:moss}
\end{subfigure}
\caption{The photographic images used as bases for the datasets in this study.}
\label{fig:base_images}
\end{figure}

\begin{figure}[ht]
\centering
\begin{subfigure}{.33\textwidth}
  \centering
  \includegraphics[width=.825\linewidth]{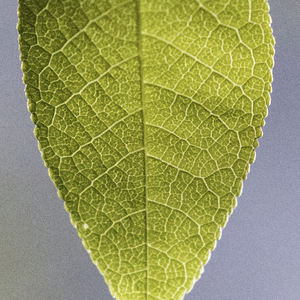}
\end{subfigure}%
\begin{subfigure}{.33\textwidth}
  \centering
  \includegraphics[width=.935\linewidth]{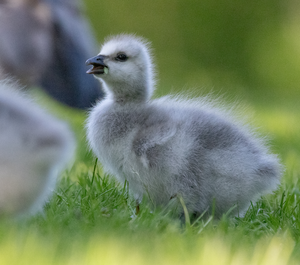}
\end{subfigure}%
\begin{subfigure}{.33\textwidth}
  \centering
  \includegraphics[width=.78\linewidth]{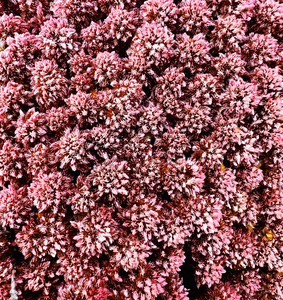}
\end{subfigure}%
\caption{The photographs from which the images in figure \ref{fig:base_images} are cropped. Images $1$ and $2$ by Markus Juvonen, $3$ by Samuli Siltanen.}
\label{fig:uncr_images}
\end{figure}

Selecting the contents of the noisy image $m$ in the model (\ref{eq:chambolle-pock}) is a crucial first step in a visual experiment. All visual data is not processed by our brain in the same way; see for example \cite{kanwisher, bianchi} and \cite{cerrahoglu}. How we interpret images is not only a visual, mechanical process, but it has a strong and largely uncharted basis in psychology and evolution as well. Emotion, memory and instinct come into play when we study images of familiar subjects. Facial recognition and pattern recognition happen subconsciously and in parts of the brain wholly dedicated to the process. Due to the highly subjective nature of perception and recognition, it is impossible to conclusively state what can and cannot be in the picture, other than that faces, face--like structures, and any emotive themes should be avoided. The images in figure \ref{fig:base_images}, which are sub--images of those in figure \ref{fig:uncr_images}, were deemed suitable for this project because they are uniform in structure and have no obvious main target.

\begin{figure}[ht]
    \centering
    \includegraphics[width=0.9\textwidth]{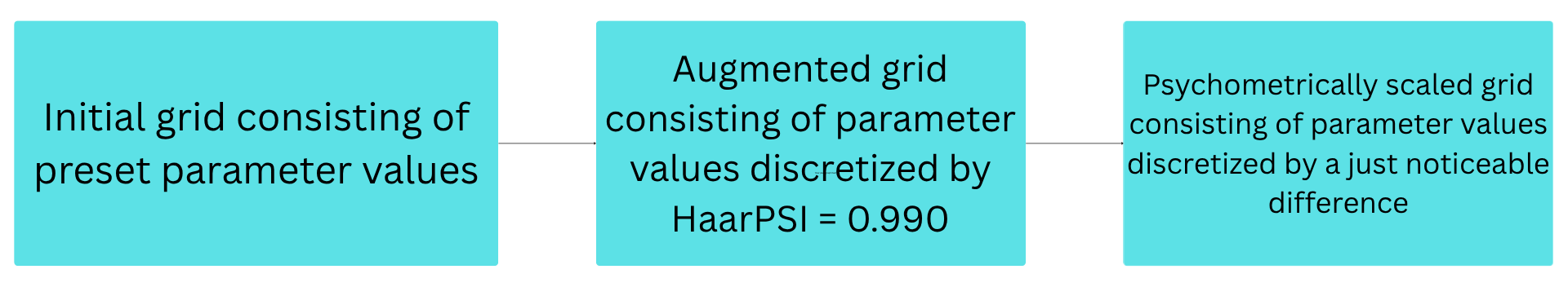}
    \caption{We start with an initial grid consisting of set values, then construct an augmented grid using it as a basis, and finally draw a psychometrically scaled grid from the augmented one. See table \ref{table:grids}.}
    \label{fig:flow}
\end{figure}

The goal of this experiment was finding a psychometrically scaled parameter grid (see figure \ref{fig:flow}). The stimulus that is needed for constructing such a grid is an augmented grid scaled by computed similarity. Given the algorithm (\ref{eq:chambolle-pock}) and noisy data $m$, a set of minimizers $f_{\alpha_j}$ is computed for $k$ regularization parameters $\alpha_j$, $j\in[1,k]$, so that any $f_{\alpha_j}$ and $f_{\alpha_{j+1}}$ have a HaarPSI (Haar--wavelet based perceptual similarity index, \cite{reisenhofer}) of at least $0.990$.

The construction of the initial grid is a rather robust process. We begin with a set
\begin{equation}
\lbrace 0.4, 2, 4, 20, 40, 200, 400, 2000, 20000 \rbrace \label{eq:init_grid}
\end{equation}
of different values of the parameter $\alpha$ in equation (\ref{eq:chambolle-pock}). 
The previously mentioned Chambolle--Pock algorithm \cite{c-p} is run for each of these values and image $m$. The set \ref{eq:init_grid} is now a starting point for an augmented parameter grid $\mathcal{G}_0$. The grid $\mathcal{G}_0$ should fulfill the criteria (i)-(ii) and (v) set for the psychometrically scaled grid $\mathcal{G}$ in the introduction. Table \ref{table:grids} demonstrates how the grids differ in size at different stages of the experiment.

\begin{table}
    \centering
\begin{tabular}{ |p{2cm}||p{2cm}|p{2cm}|p{3cm}|  }
 \hline
 \multicolumn{4}{|c|}{The size of each grid} \\
 \hline
 Image number & Initial grid & Augmented grid & Psychometrically scaled grid \\
 \hline
 1 & 9 & 23 &   15 \\
 2 & 9 & 32 & N/A (32)\\
 3 & 9 & 22 &  8\\
 \hline
\end{tabular}
\caption{A summary of the set sizes.}
\label{table:grids}
\end{table}

We consider the grid $\mathcal{G}_0$ to have a structure of an ordered set with elements arranged in ascending order by value, so that the first element ($\alpha_1$) is the smallest parameter, and the last element ($\alpha_k$) is the largest parameter. These parameters fulfill criteria (i)-(ii). If $\alpha_m, \alpha_n \in \mathcal{G}_0$, $\alpha_m < \alpha_n$, and there is no such $\alpha_p \in \mathcal{G}_0$ that $\alpha_m < \alpha_p < \alpha_n$, then $\alpha_m$ and $\alpha_n$ are \emph{consecutive} parameters. Accordingly, we call the solutions $f_{\alpha_m}$ and $f_{\alpha_n}$ consecutive images.

\begin{figure}[ht]
\centering
\begin{subfigure}{.3\textwidth}
  \centering
  \includegraphics[width=.9\linewidth]{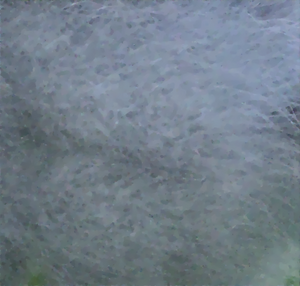}
\end{subfigure}%
\begin{subfigure}{.3\textwidth}
  \centering
  \includegraphics[width=.9\linewidth]{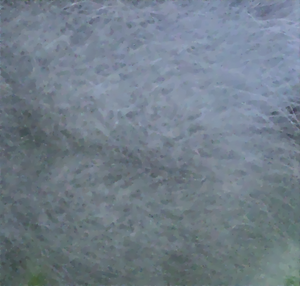}
\end{subfigure}
\caption{Two consecutive samples from set $2$. The images $23$ and $24$ have a HaarPSI of $0.996$.}
\label{fig:consec_im}
\end{figure}

The grid $\mathcal{G}_0$ can be built by filling the set (\ref{eq:init_grid}) in with parameter values so that consecutive images are equally spaced in terms of similarity index value, from which criterion (iii) follows. We want the augmented grid to be quite fine, and as a result of the experiment, to be able to draw out a smaller subset of images from the sample set; thus, fulfilling criterion (v). The value was selected visually to be higher than $0.990$ (see figure \ref{fig:consec_im}). Here, whenever consecutive images $f_{\alpha_a}$ and $f_{\alpha_b}$ have a HaarPSI value smaller than $0.990$, a value $\alpha_c$ is added to the set (\ref{eq:init_grid}) so that $\alpha_a < \alpha_c < \alpha_b$. 
Then, a new TV--denoised image is computed using $\alpha_c$ as the parameter, and similarity indices are computed for the pairs $f_{\alpha_a}$ and $f_{\alpha_c}$, and $f_{\alpha_b}$ and $f_{\alpha_c}$. The resulting grid is
\begin{equation}
    \mathcal{G}_0 = \lbrace \alpha_1, \alpha_2, \dots , \alpha_k\rbrace \label{eq:sample_grid}
\end{equation}
and the sample set is
\begin{equation}
    \mathcal{F} = \lbrace f_{\alpha_1}, f_{\alpha_2}, \dots , f_{\alpha_k}\rbrace \label{eq:sample_set} .
\end{equation}

Once the sample set is complete, a comparison test is conducted. Comparing images that have been constructed using significantly different parameters is not interesting or relevant to this study. As we are interested in finding a threshold for a just noticeable difference, we can limit the sample pairs to those that are quite similar to each other. If the first image drawn for comparison is $f_{\alpha_p}$ in the set $\mathcal{F}$, the second image $f_{\alpha_q}$ is such that $|p-q| \leq 4$. 
We decided to not include pairs consisting of the same image twice, as it does not bring additional value to the research. Without these limitations, the amount of all possible pairs to evaluate in the experiment would be $506$ for a set of $23$ samples, whereas with them, the amount is only $164$.

\begin{figure}[ht]
\centering
\begin{subfigure}{.3\textwidth}
  \centering
  \includegraphics[width=.9\linewidth]{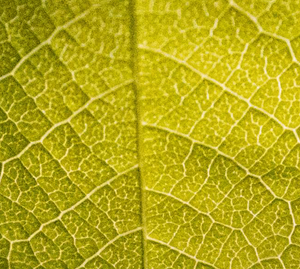}
\end{subfigure}%
\begin{subfigure}{.3\textwidth}
  \centering
  \includegraphics[width=.9\linewidth]{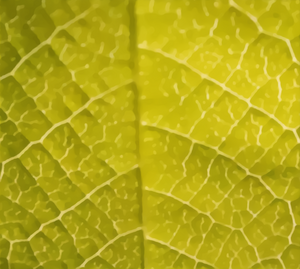}
\end{subfigure}
\caption{Images $1$ and $23$ of set $1$.}
\label{fig:leaf_endpoints}
\end{figure}
\begin{figure}[ht]
\centering
\begin{subfigure}{.3\textwidth}
  \centering
  \includegraphics[width=.9\linewidth]{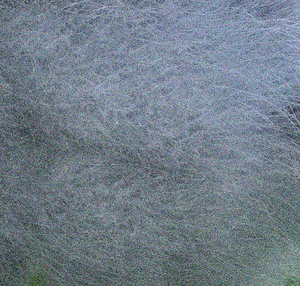}
\end{subfigure}%
\begin{subfigure}{.3\textwidth}
  \centering
  \includegraphics[width=.9\linewidth]{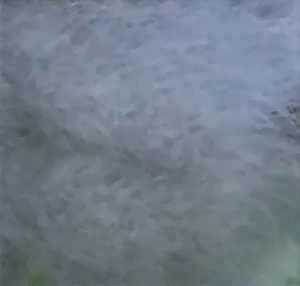}
\end{subfigure}
\caption{Images $1$ and $32$ of set $2$.}
\label{fig:down_endpoints}
\end{figure}
\begin{figure}[ht]
\centering
\begin{subfigure}{.3\textwidth}
  \centering
  \includegraphics[width=.9\linewidth]{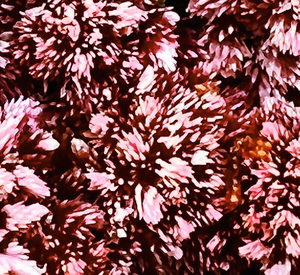}
\end{subfigure}%
\begin{subfigure}{.3\textwidth}
  \centering
  \includegraphics[width=.9\linewidth]{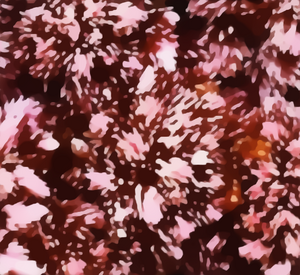}
\end{subfigure}
\caption{Images $1$ and $22$ of set $3$.}
\label{fig:moss_endpoints}
\end{figure}
\begin{figure}[p]
\begin{picture}(300,500)
    \put(0,500){\includegraphics[width=78.56pt]{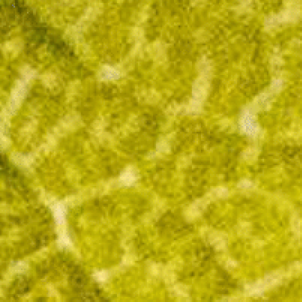}}
    \put(100,500){\includegraphics[width=78.56pt]{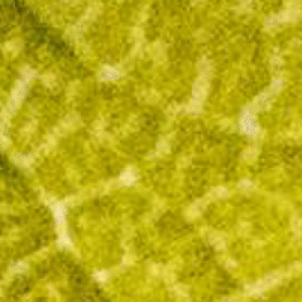}}
    \put(200,500){\includegraphics[width=78.56pt]{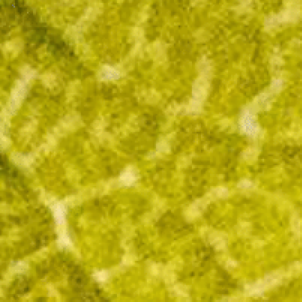}}
    \put(300,500){\includegraphics[width=78.56pt]{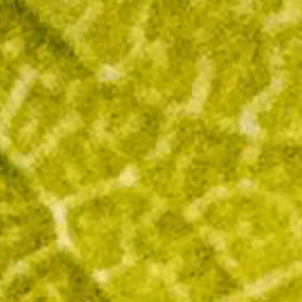}}
    \put(0,400){\includegraphics[width=78.56pt]{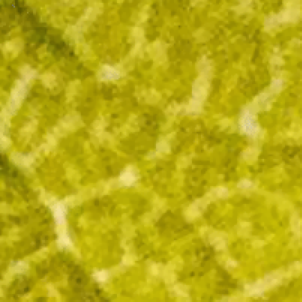}}
    \put(100,400){\includegraphics[width=78.56pt]{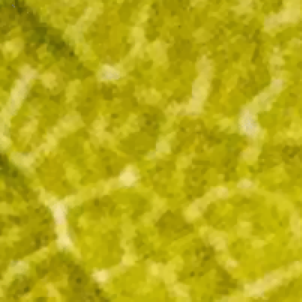}}
    \put(200,400){\includegraphics[width=78.56pt]{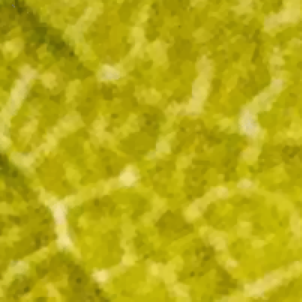}}
    \put(300,400){\includegraphics[width=78.56pt]{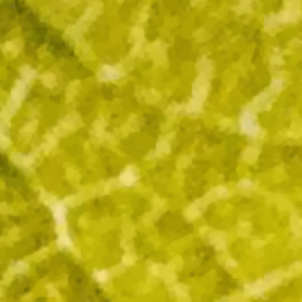}}
    \put(0,300){\includegraphics[width=78.56pt]{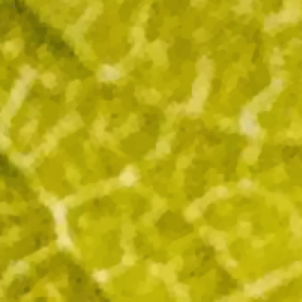}}
    \put(100,300){\includegraphics[width=78.56pt]{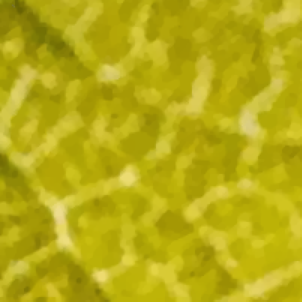}}
    \put(200,300){\includegraphics[width=78.56pt]{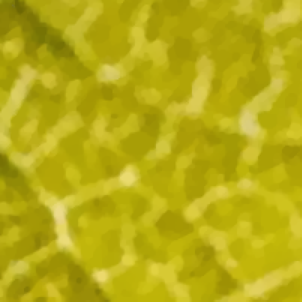}}
    \put(300,300){\includegraphics[width=78.56pt]{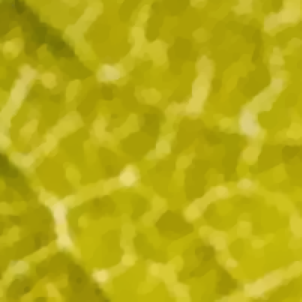}}
    \put(0,200){\includegraphics[width=78.56pt]{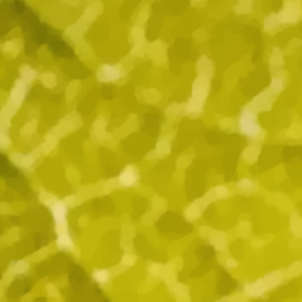}}
    \put(100,200){\includegraphics[width=78.56pt]{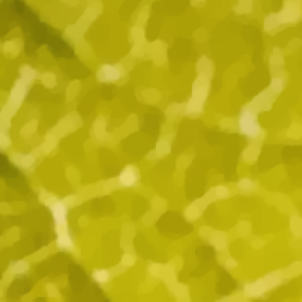}}
    \put(200,200){\includegraphics[width=78.56pt]{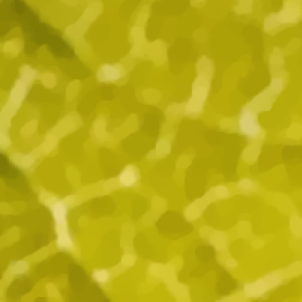}}
    \put(300,200){\includegraphics[width=78.56pt]{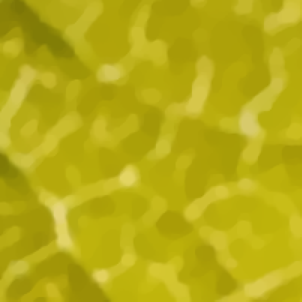}}
    \put(0,100){\includegraphics[width=78.56pt]{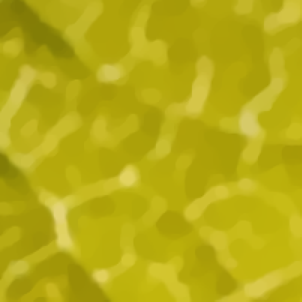}}
    \put(100,100){\includegraphics[width=78.56pt]{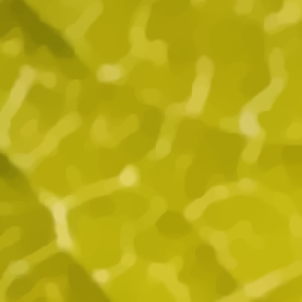}}
    \put(200,100){\includegraphics[width=78.56pt]{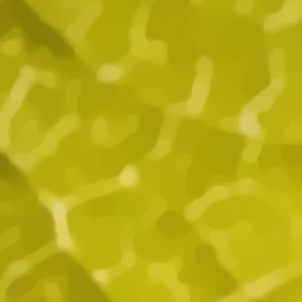}}
    \put(300,100){\includegraphics[width=78.56pt]{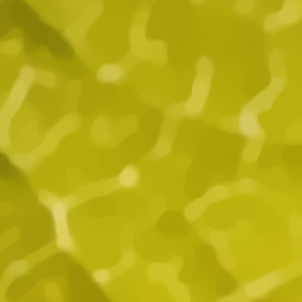}}
    \put(50,0){\includegraphics[width=78.56pt]{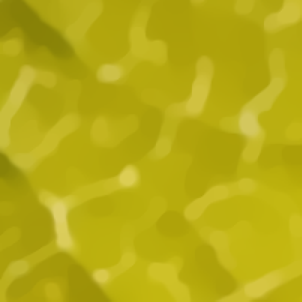}}
    \put(150,0){\includegraphics[width=78.56pt]{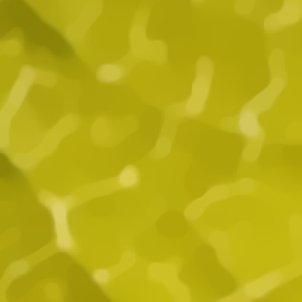}}
    \put(250,0){\includegraphics[width=78.56pt]{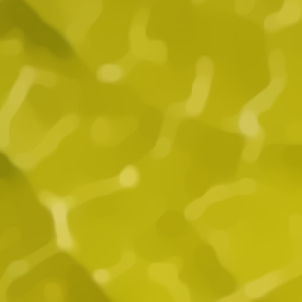}}
\end{picture}
\caption{\label{fig:leaf_set} A smaller section of each image in set $1$ has been magnified here to better display the different noise levels brought on by different values of the parameter $\alpha$. The full-sized images are available on Zenodo.}
\end{figure}
\begin{figure}[p]
\begin{picture}(300,420)
    \put(0,420){\includegraphics[width=60.7pt]{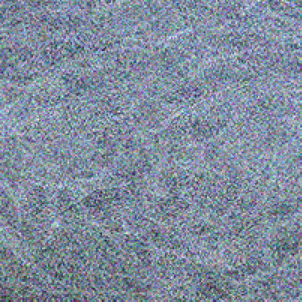}}
    \put(75,420){\includegraphics[width=60.7pt]{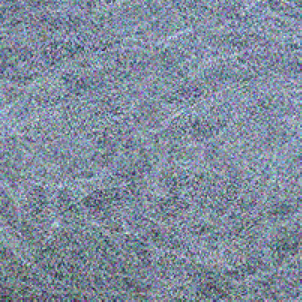}}
    \put(150,420){\includegraphics[width=60.7pt]{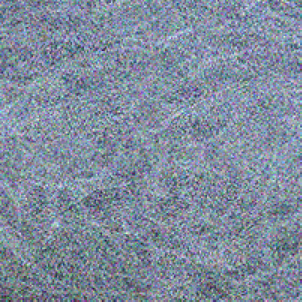}}
    \put(225,420){\includegraphics[width=60.7pt]{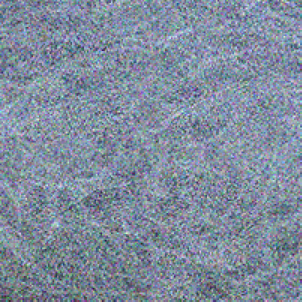}}
    \put(300,420){\includegraphics[width=60.7pt]{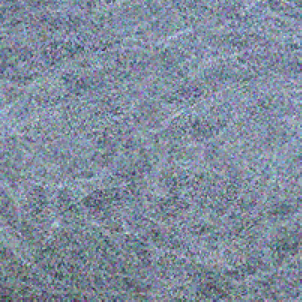}}
    \put(0,350){\includegraphics[width=60.7pt]{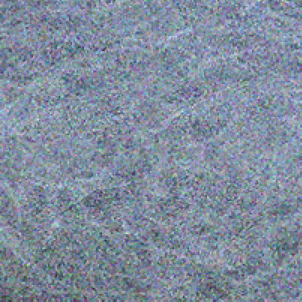}}
    \put(75,350){\includegraphics[width=60.7pt]{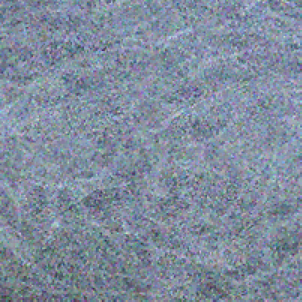}}
    \put(150,350){\includegraphics[width=60.7pt]{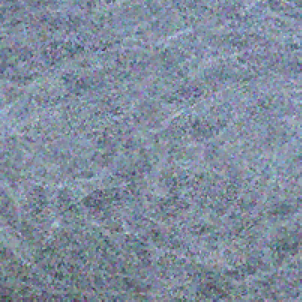}}
    \put(225,350){\includegraphics[width=60.7pt]{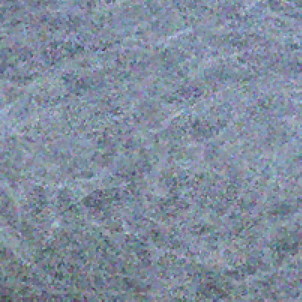}}
    \put(300,350){\includegraphics[width=60.7pt]{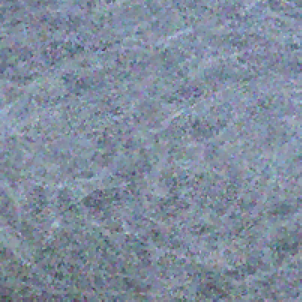}}
    \put(0,280){\includegraphics[width=60.7pt]{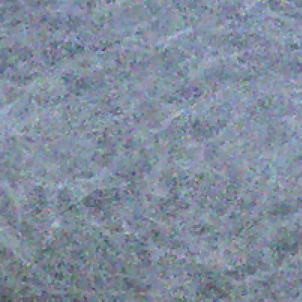}}
    \put(75,280){\includegraphics[width=60.7pt]{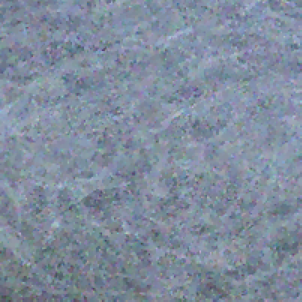}}
    \put(150,280){\includegraphics[width=60.7pt]{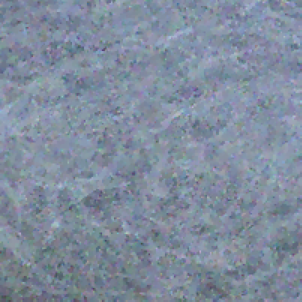}}
    \put(225,280){\includegraphics[width=60.7pt]{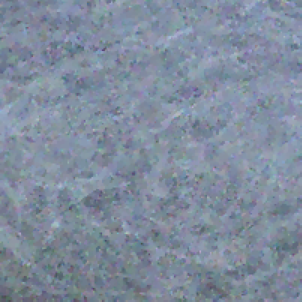}}
    \put(300,280){\includegraphics[width=60.7pt]{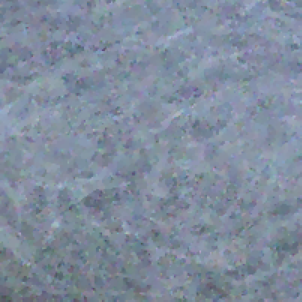}}
    \put(0,210){\includegraphics[width=60.7pt]{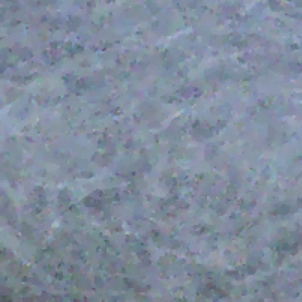}}
    \put(75,210){\includegraphics[width=60.7pt]{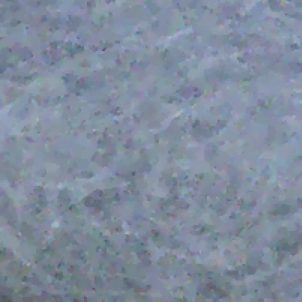}}
    \put(150,210){\includegraphics[width=60.7pt]{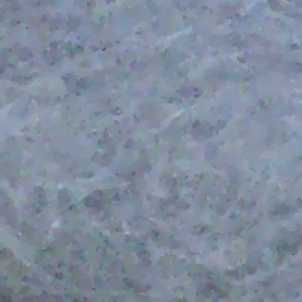}}
    \put(225,210){\includegraphics[width=60.7pt]{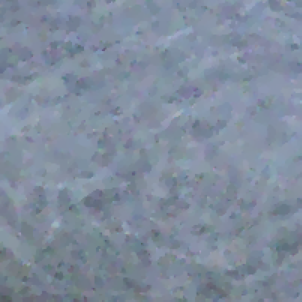}}
    \put(300,210){\includegraphics[width=60.7pt]{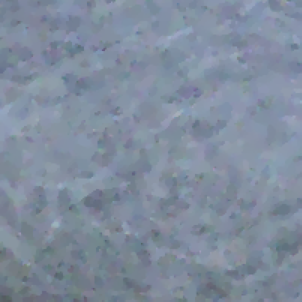}}
    \put(0,140){\includegraphics[width=60.7pt]{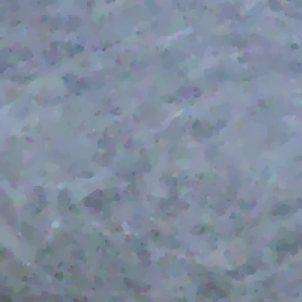}}
    \put(75,140){\includegraphics[width=60.7pt]{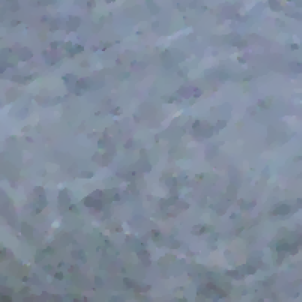}}
    \put(150,140){\includegraphics[width=60.7pt]{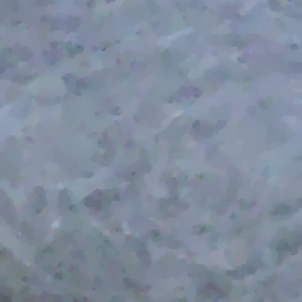}}
    \put(225,140){\includegraphics[width=60.7pt]{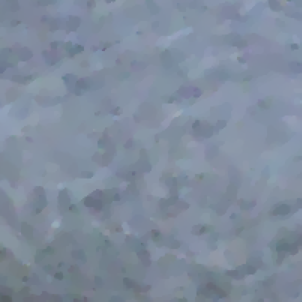}}
    \put(300,140){\includegraphics[width=60.7pt]{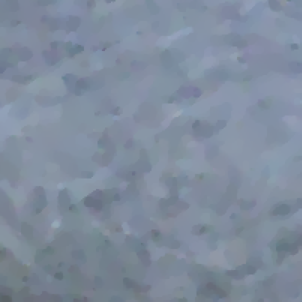}}
    \put(0,70){\includegraphics[width=60.7pt]{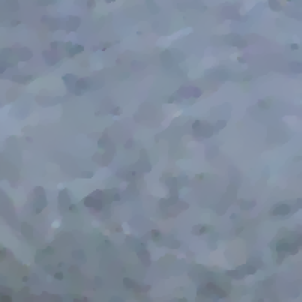}}
    \put(75,70){\includegraphics[width=60.7pt]{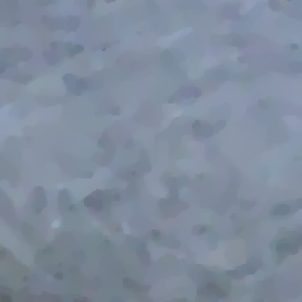}}
    \put(150,70){\includegraphics[width=60.7pt]{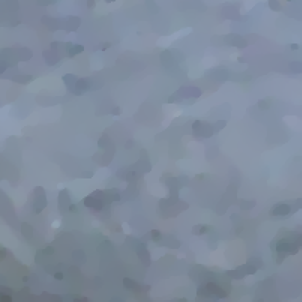}}
    \put(225,70){\includegraphics[width=60.7pt]{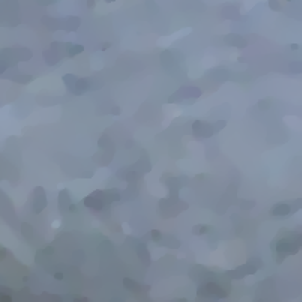}}
    \put(300,70){\includegraphics[width=60.7pt]{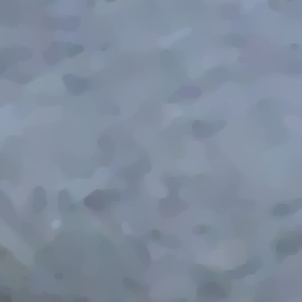}}
    \put(113,0){\includegraphics[width=60.7pt]{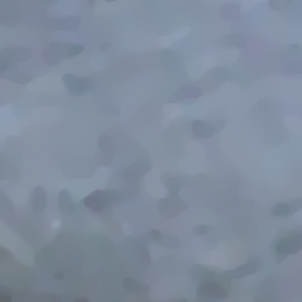}}
    \put(188,0){\includegraphics[width=60.7pt]{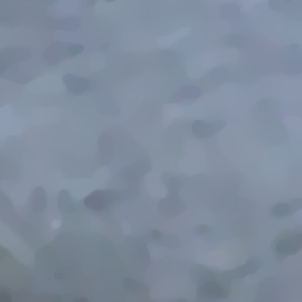}}
\end{picture}
\caption{\label{fig:down_set} The magnified sub--images of set $2$.}
\end{figure}
\begin{figure}[p]
\begin{picture}(300,500)
    \put(0,500){\includegraphics[width=82.05pt]{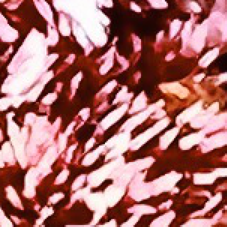}}
    \put(100,500){\includegraphics[width=82.05pt]{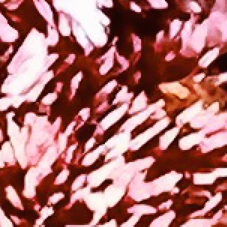}}
    \put(200,500){\includegraphics[width=82.05pt]{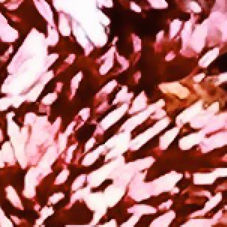}}
    \put(300,500){\includegraphics[width=82.05pt]{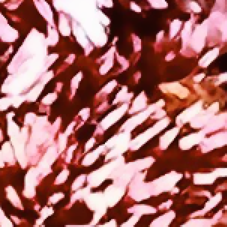}}
    \put(0,400){\includegraphics[width=82.05pt]{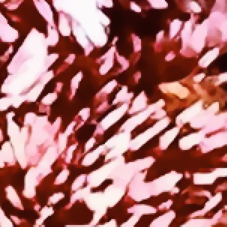}}
    \put(100,400){\includegraphics[width=82.05pt]{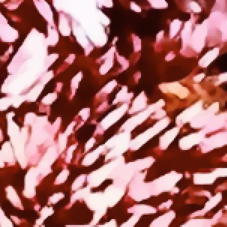}}
    \put(200,400){\includegraphics[width=82.05pt]{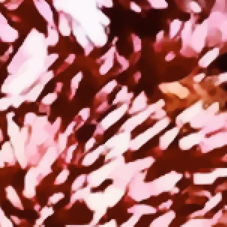}}
    \put(300,400){\includegraphics[width=82.05pt]{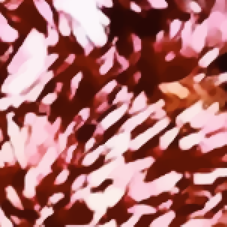}}
    \put(0,300){\includegraphics[width=82.05pt]{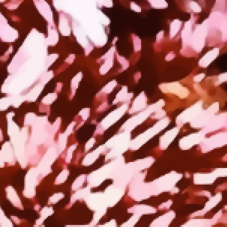}}
    \put(100,300){\includegraphics[width=82.05pt]{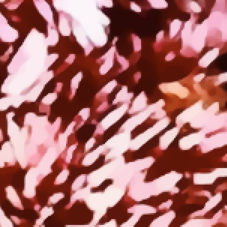}}
    \put(200,300){\includegraphics[width=82.05pt]{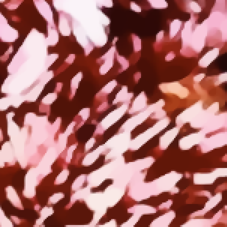}}
    \put(300,300){\includegraphics[width=82.05pt]{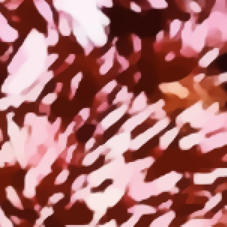}}
    \put(0,200){\includegraphics[width=82.05pt]{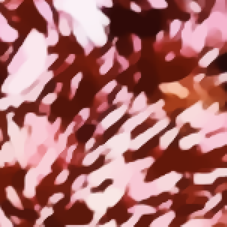}}
    \put(100,200){\includegraphics[width=82.05pt]{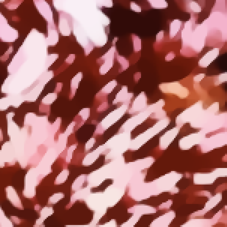}}
    \put(200,200){\includegraphics[width=82.05pt]{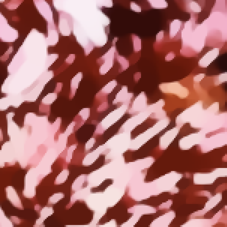}}
    \put(300,200){\includegraphics[width=82.05pt]{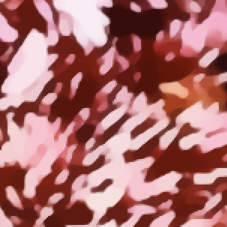}}
    \put(0,100){\includegraphics[width=82.05pt]{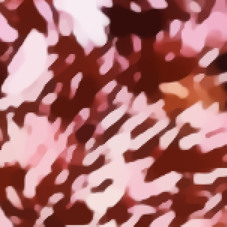}}
    \put(100,100){\includegraphics[width=82.05pt]{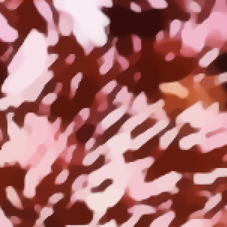}}
    \put(200,100){\includegraphics[width=82.05pt]{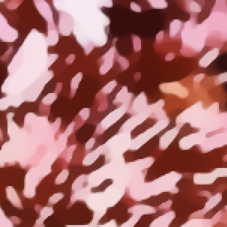}}
    \put(300,100){\includegraphics[width=82.05pt]{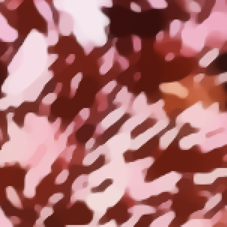}}
    \put(100,0){\includegraphics[width=82.05pt]{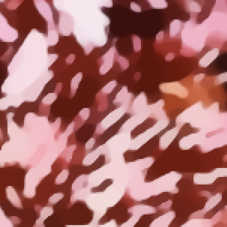}}
    \put(200,0){\includegraphics[width=82.05pt]{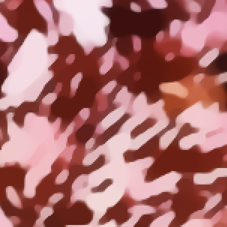}}
\end{picture}
\caption{\label{fig:moss_set} The magnified sub--images of set $3$.}
\end{figure}

In the remainder of this paper, "image set $1$", "$2$", or "$3$" refers to the augmented set constructed using the corresponding image $1$, $2$, or $3$ (figure \ref{fig:base_images}) as a basis. The noisiest and smoothest image in each set, corresponding to the smallest and the largest value of $\alpha$ respectively, can be seen in figures \ref{fig:leaf_endpoints}, \ref{fig:down_endpoints} and \ref{fig:moss_endpoints}. The complete sets can be found in full size on Zenodo \footnote{\url{https://doi.org/10.5281/zenodo.18457707}} \cite{zenodo}, and in smaller size in figures \ref{fig:leaf_set}, \ref{fig:down_set} and \ref{fig:moss_set}. In those figures a sub-image of each separate image is displayed to visually accentuate the differences between them. Within each set, the images were labeled according to regularization parameter in ascending order. Set $1$ consists of $23$ images of pixel size $982\times 882$ ($F = \lbrace f_{\alpha_1}, \dots , f_{\alpha_{23}} \rbrace$), set $2$ of $32$ images of pixel size $607 \times 579$ ($G = \lbrace g_{\beta_1}, \dots , g_{\beta_{32}} \rbrace$), and set $3$ of $22$ images of pixel size $547 \times 501$ ($H = \lbrace h_{\gamma_1}, \dots , h_{\gamma_{22}} \rbrace$).

The test was conducted online \url{https://compare.blasten.eu} \footnote{Available at the time of writing this article} \cite{site}, built and administrated by our research team. 
The image pairs were displayed side by side on the same screen as in figure \ref{fig:site}. In the in--person setting participants viewed them on a Lenovo 21L2S0V400 $14$'' LCD display. In remote participation observers were instructed to use a computer screen or a laptop. The comparisons made by the same user and at what time are visible in the data, but beyond that the user data is anonymous.

\subsection{Procedure} \label{chapter: procedure}

Focusing on only one scalable quality, and gathering data in a binary form allows us to research perception rather rigorously despite the complexity of the process. 
Comparison tests, or $A/B$--tests, all have the same basis. Observers view samples in pairs, threes, or one by one, and make an evaluation of each viewing. The experiment in this project is based on \emph{the method of constant stimuli} (chapter $5$, \cite{engeldrum}). An observer is shown a reference image and a set of samples one by one, and then asked if the sample contains the quality the researchers are interested in. In this case, the observer is shown two images denoised with different values of the parameter $\alpha$, and asked if they are equally noisy.

In the method of constant stimuli the binary data consisting of ``yes''- or ``no''-answers is translated to scalar values as proportions of ``yes'' -answers. With the range of these values on the $y$--axis, they can be plotted against the samples on the $x$--axis. We can then fit a continuous graph to these points, forming the \emph{psychometric curve}, which treats the level of noise in an image as a measure of its quality compared to similar images. JND and uncertainty intervals can be evaluated directly from the psychometric curve.

\begin{figure}[ht]
    \centering
    \includegraphics[width=0.9\textwidth]{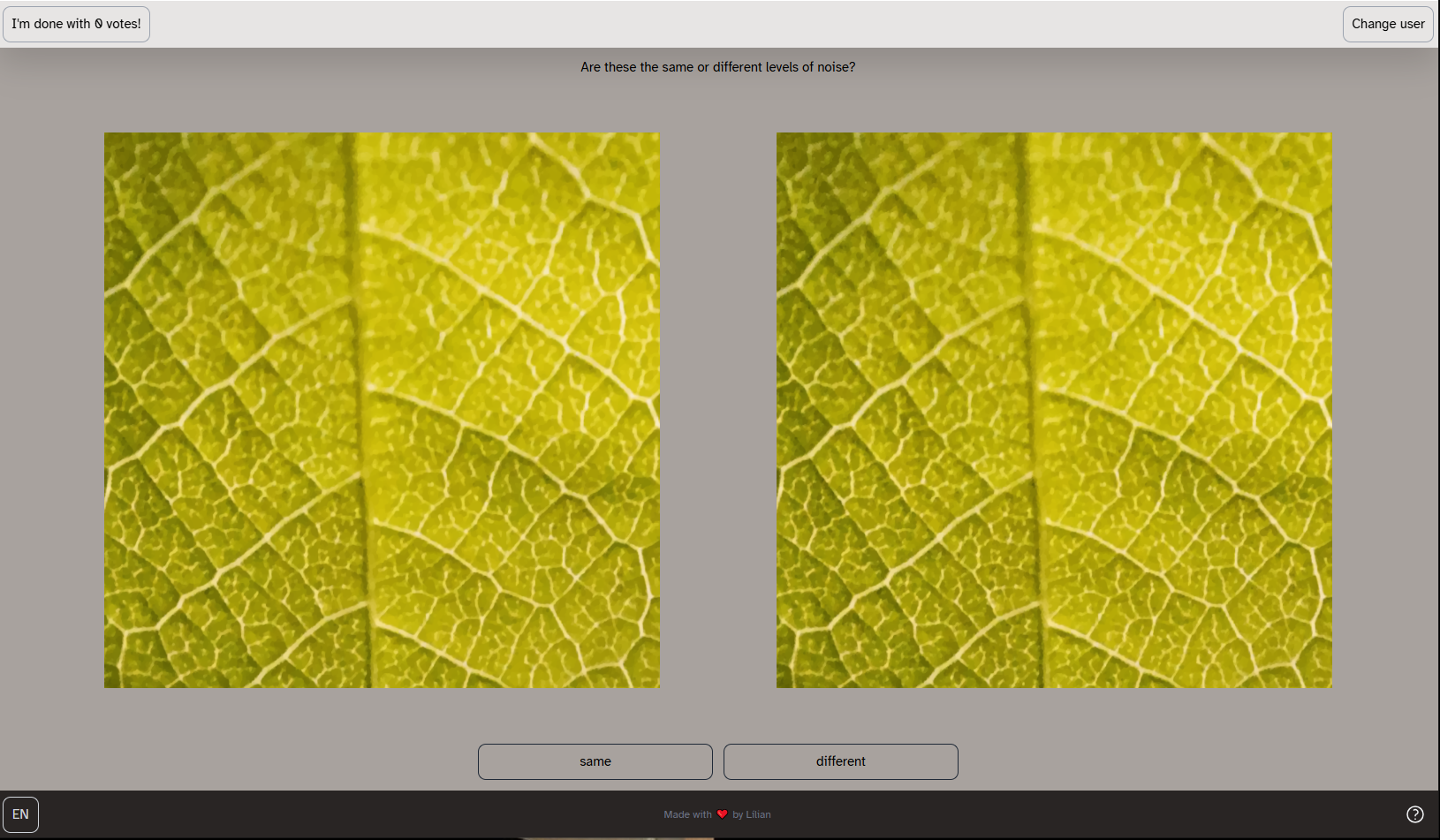}
    \caption{\emph{A screen capture of the test environment \cite{site}.}}
    \label{fig:site}
\end{figure}

The experiment must be designed to produce results that are as reliable as possible. Thus, the amount of true, neutral evaluations of the sample data must be maximized. Two major factors that contribute to this are, first of all, the maximum amount of evaluations that can be made in one sitting, and, second of all, if the instructions are clear enough. Four separate rounds of experiments were conducted, and participants provided feedback throughout the project. The test setting was adjusted as needed between rounds. An example screen capture of the test display is shown in figure \ref{fig:site}.

In experiments like these, comfort of the participants has to be taken into account, as well as the explicit issue of growing bored. After a certain point in a test with an infinite amount of questions, observers lose their interest in the images which greatly affects their answers. 
The effects of boredom (in behavioral studies) have been researched quite recently for example by Maria Meier et al. \cite{meier}.

The question, ``Are these the same or different levels of noise?", was posed to be as clear and short as possible, so that the observer would know immediately what they would have to do, and so that interpreting the question would not draw their focus away from evaluating the images. The question also has to be posed without implication of a right or wrong answer. However, to people without a background in mathematics the term ``noise'' might not be clear. The team still felt that there was no synonymous, popular word to replace it with without losing some significance ("quality", for example, might be too ambiguous). The wording of the question was then kept as is, but the decision was made to add an \emph{anchor image} (such as figure \ref{fig:anchor}) to the beginning of the test to demonstrate the effects of noise in imaging.

The background color of the experiment site was changed from dark gray to a lighter one after the second block. It appeared that the contrast between the background and the images in the original layout cast a faint illusion that the left--hand--side image is darker than the right one. A comment that was often heard during the second round was that it was ``more difficult'' than the first one. We can see in the results a lower threshold value for JND of the noisy images of block $2$. An observation to be made from this is that participants are likely to feel a sense of accomplishment or victory when they find differences between images, similar to a puzzle. The possible tendency to hope that the images are different should be accounted for in further experiments, perhaps in the posing of the research question or participant instruction.

The experiment consisted of four blocks. In order to ensure that a sufficient amount of evaluations is recorded per image pair, the amount of sample data was limited in each block. See table \ref{table:data} for a summary. In the first block, images $f_{\alpha_{16}}$ to $f_{\alpha_{23}}$ in image set $1$ were surveyed. In this block, all participants answered the survey with a member of the research team supervising in the same room. The second block was conducted using images $f_{\alpha_1}$ to $f_{\alpha_8}$ in image set $1$, with one group of participants answering the survey in similar settings as the first round, and another group answering it independently online on their own devices.

The third block contained images $f_{\alpha_{10}}$ to $f_{\alpha_{14}}$ and $g_{\beta_{11}}$ to $g_{\beta_{17}}$ from sets $1$ and $2$ (the "middle values") and was executed completely remotely. Finally, the fourth block was conducted using images $h_{\gamma{10}}$ to $h_{\gamma_{19}}$ in set $3$. As well as being available online, the experiment was displayed in the Researchers' night event \footnote{see \url{https://tutkijoidenyo.fi/en/} and \url{https://www.helsinki.fi/en/tiedekasvatus/researchers-night}} in Helsinki in September $2025$, where volunteers from the general public participated in the survey \cite{researchers_night}. The anchor image \ref{fig:anchor} was introduced at this point.

The comparison test was conducted as multiple concurrent experiments following the method of constant stimuli \cite{engeldrum}. The participant - or ``user'' - sees a pair of images and is asked to evaluate whether they contain same or different levels of noise. The selection is made by clicking a button at the bottom of the page (see figure \ref{fig:site}). The data obtained this way consists of information regarding which image pairs were shown and how the participant evaluated them. Users were recommended to evaluate approximately $50$ samples, but to stop if they began to feel bored. The participants were told that there is no time limit for answering the questions, and that there are no right or wrong answers.

$565$ evaluations were given in block $1$. The median value for evaluations provided per participant was $50$. $547$ evaluations were given in block $2$ with a median of $18.5$. Block $3$ resulted in $398$ evaluations with $50$ for the median value, and block $4$ had $449$ evaluations and a median value of $16$. The total amount of evaluations in the experiment was $1959$, with a mean value for evaluations provided per participant at $28$ evaluations, and a median at $25$ evaluations. Table \ref{table:data} displays a conclusion of the data.

\bigskip

\subsection{Analysis} \label{chapter:analysis}


Chapter \ref{chapter:stimuli} addressed the construction of parameter scales discretized by computed similarity. We resume the discussion here, and describe how the observer--based data obtained in the experiment can be used to re--discretize the augmented grid.

Once the discrete data obtained per each sample pair is plotted using Matlab, a graph is fitted to it (see figures \ref{fig:1d_leaf} and \ref{fig:2and3}). This is the psychometric curve constructed from the dataset in question. The curve yields us a threshold value for a just noticeable difference. The JND threshold is a point on the $x$--axis, found at the point where the curve and the line $y = 0.5$ intersect. The newly found threshold is a HaarPSI value - $x= x_{JND}$ - by which we can re--discretize the grid (\ref{eq:sample_grid}). The elements of the psychometrically scaled grid $\mathcal{G}$ (\ref{eq:grid_from_intro}) will be selected from the augmented parameter grid, $\mathcal{G}_0$, so that $\mathcal{G} \subseteq \mathcal{G}_0$. The images denoised with these parameters now form the set
\begin{equation}
    \mathcal{F}_{\mathcal{G}} = \lbrace f_{\alpha^*_1}, f_{\alpha^*_2}, \dots , f_{\alpha^*_K} \rbrace. \label{eq:psych_set}
\end{equation}
This discretization results in a psychometrically scaled parameter grid and the associated image set that meets the criteria (iv)-(v) listed in the introduction, which is the goal of this scaling study.

The set $\mathcal{G}$ is the smallest possible set so that any consecutive images in $\mathcal{F}_{\mathcal{G}}$ will have a HaarPSI of at least $x_{JND}$.

\subsection{Results} \label{chapter:results}

\begin{figure}[ht]
    \centering
    \includegraphics[width=0.6\textwidth]{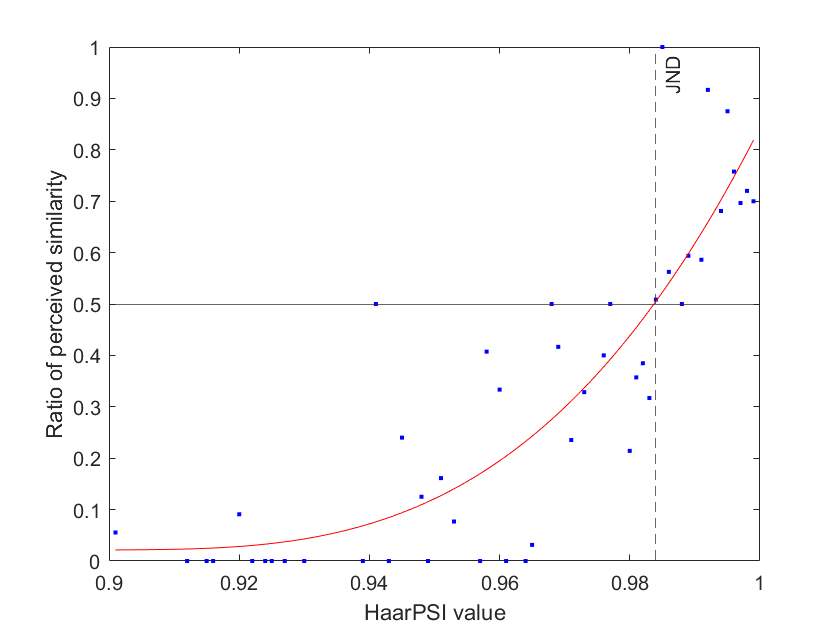}
    \caption{The psychometric curve of set $1$. Every blue dot in the image is related to one specific image pair in the data set. The $x$-coordinate of a blue dot is the HaarPSI similarity between the two images. The $y$-coordinate gives the proportion of positive answers by participants to the question ``Are these the same or different levels of noise?''}
    \label{fig:1d_leaf}
\end{figure}

The psychometric curve for image set $1$ can be seen in figure \ref{fig:1d_leaf}, with the just noticeable difference being $0.984$. The $x$--axis contains the HaarPSI values of the image pairs in the sample set, and the $y$--axis depicts a percentage value between $0$ and $1$. The discrete points mark the proportion of "yes"--answers associated with the similarity indices. We can see, for instance, that two images having a HaarPSI of approximately $0.92$ were evaluated as similar by approximately $10$ per cent of observers, or that approximately $60$ per cent of observers evaluated images having a HaarPSI of approximately $0.99$ to be similar. The curve is the best fit to model the correlation between those evaluations and the similarity indices. With it, we can predict and estimate how the strength of denoising affects how the image is perceived.

\begin{figure}[ht]
\centering
\begin{subfigure}{.5\textwidth}
  \centering
  \includegraphics[width=.9\linewidth]{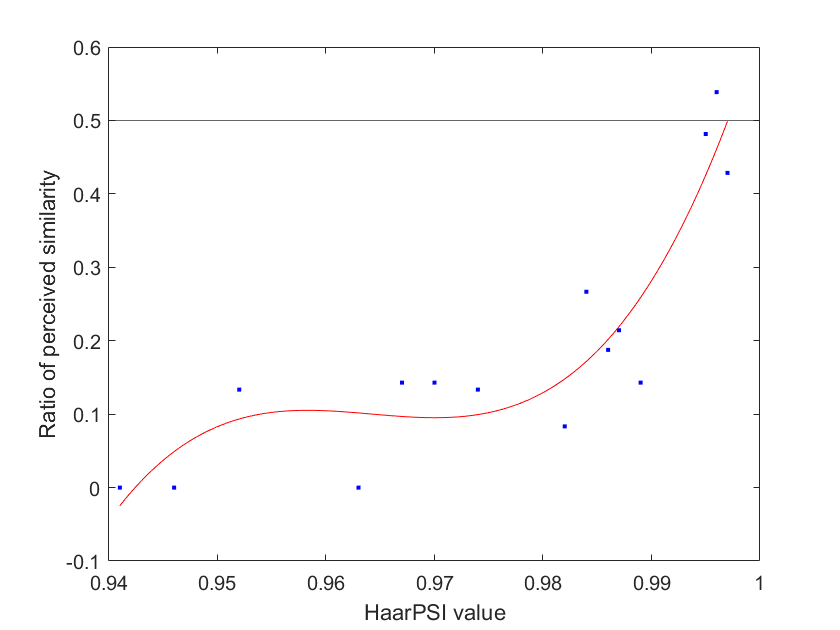}
  \caption{\emph{Set $2$}}
  \label{fig:1d_down}
\end{subfigure}%
\begin{subfigure}{.5\textwidth}
  \centering
  \includegraphics[width=.9\linewidth]{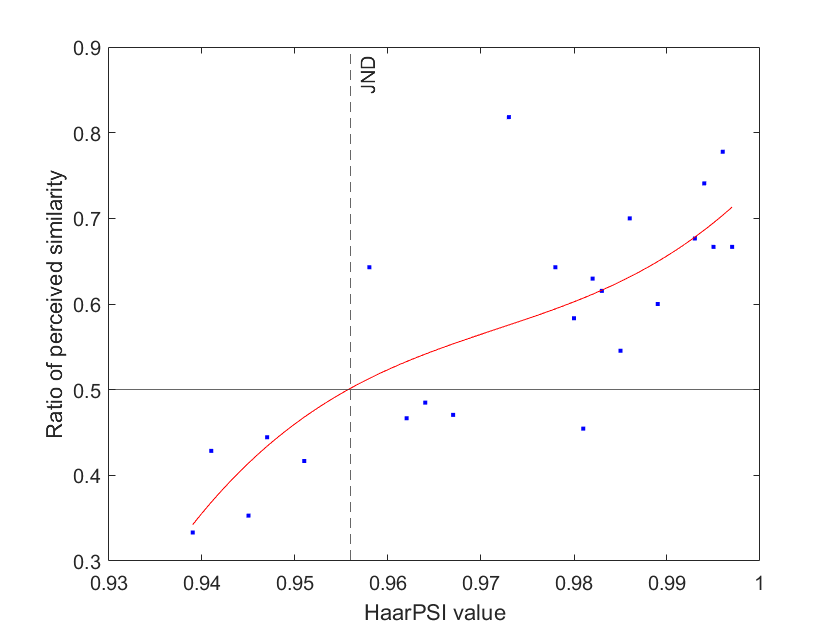}
  \caption{\emph{Set $3$}}
  \label{fig:1d_moss}
\end{subfigure}
\caption{The psychometric curves of sets $2$ and $3$.}
\label{fig:2and3}
\end{figure}

The psychometric curve constructed from the data obtained in block $3$ for image $2$ can be seen in figure \ref{fig:1d_down}. Interestingly, the entire curve lies below the value $y=0.5$, meaning that we cannot define a threshold for a just noticeable difference for this image, at least using this dataset as a basis. The threshold value for this image is likely higher than $0.990$, and should be surveyed using a finer parameter grid than the sample grid defined here. Then again, this data implies that the images in this set are already perceptually different from each other. The curve also appears to just intersect with $y=0.5$, which can be interpreted to mean that the original scaling coincides with perceived similarity. Keeping in mind the possibility that some information may be missing, the set could be used for further studies in perceived noisiness as is.

Figure \ref{fig:1d_moss} displays the psychometric curve constructed from the data obtained in experiment $4$ for image $3$. The threshold value found for a just noticeable difference based on this data is as low as $0.956$.

\begin{figure}[p]
\begin{picture}(320, 260)(-30, 0)
    \put(-30,260){\includegraphics[width=78.56pt]{images/leaf2_23.png}}
    \put(60,260){\includegraphics[width=78.56pt]{images/leaf2_20.png}}
    \put(150,260){\includegraphics[width=78.56pt]{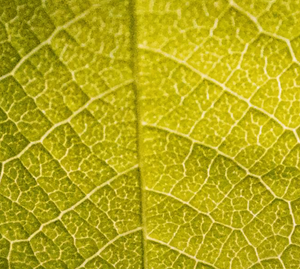}}
    \put(240,260){\includegraphics[width=78.56pt]{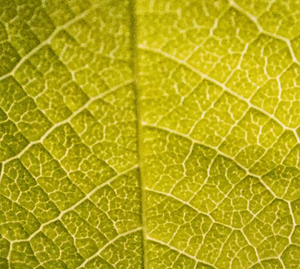}}
    \put(-30,180){\includegraphics[width=78.56pt]{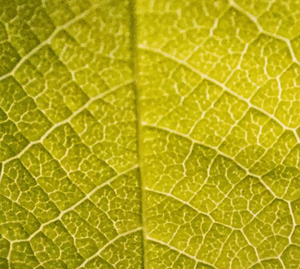}}
    \put(60,180){\includegraphics[width=78.56pt]{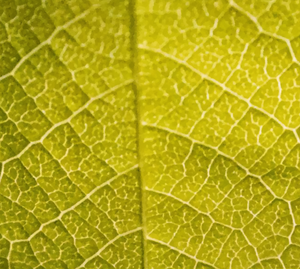}}
    \put(150,180){\includegraphics[width=78.56pt]{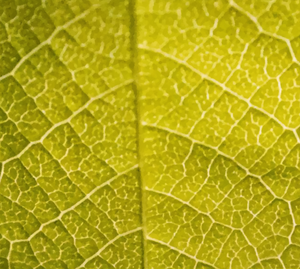}}
    \put(240,180){\includegraphics[width=78.56pt]{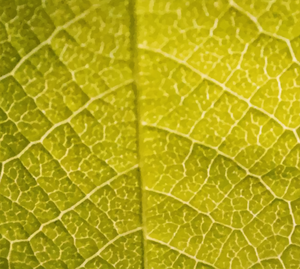}}
    \put(-30,100){\includegraphics[width=78.56pt]{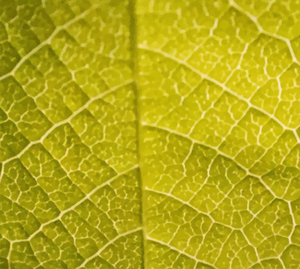}}
    \put(60,100){\includegraphics[width=78.56pt]{images/leaf2_08.png}}
    \put(150,100){\includegraphics[width=78.56pt]{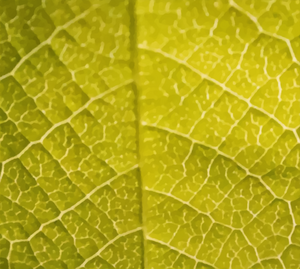}}
    \put(240,100){\includegraphics[width=78.56pt]{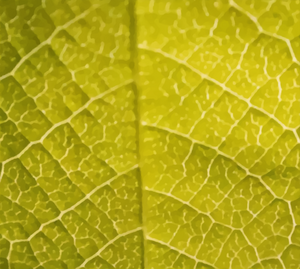}}
    \put(15,20){\includegraphics[width=78.56pt]{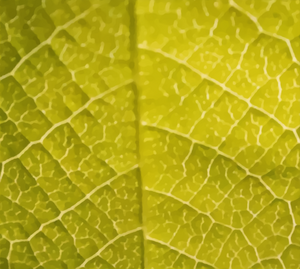}}
    \put(105,20){\includegraphics[width=78.56pt]{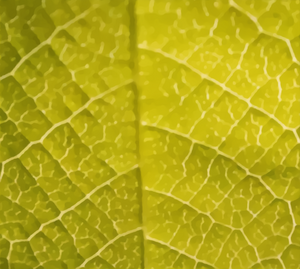}}
    \put(195,20){\includegraphics[width=78.56pt]{images/leaf2_01.png}}
\end{picture}
\caption{\label{fig:leaf_psych_set} Psychometrically scaled set $1$.}
\end{figure}

\begin{figure}

\begin{picture}(320,170)(-30,0)
\put(-30,120){\includegraphics[width=3cm]{images/terrain2_22.png}}
\put(60,120){\includegraphics[width=3cm]{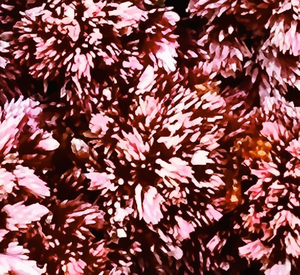}}
\put(150,120){\includegraphics[width=3cm]{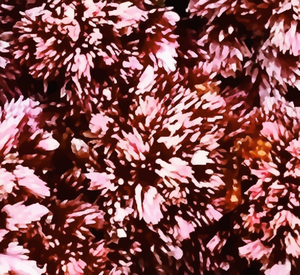}}
\put(240,120){\includegraphics[width=3cm]{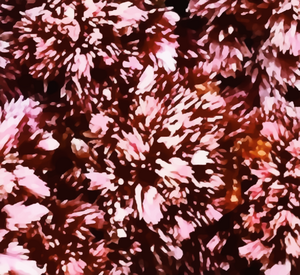}}
\put(-30,30){\includegraphics[width=3cm]{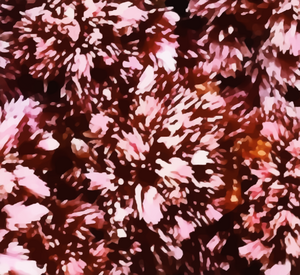}}
\put(60,30){\includegraphics[width=3cm]{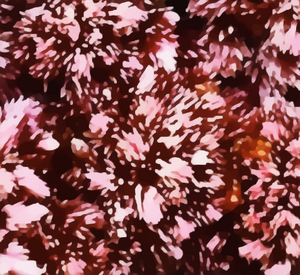}}
\put(150,30){\includegraphics[width=3cm]{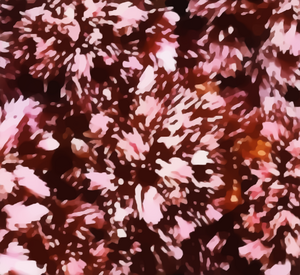}}
\put(240,30){\includegraphics[width=3cm]{images/terrain2_01.png}}
\end{picture}
\caption{\label{fig:moss_psych_set} Psychometrically scaled set $3$.}
\end{figure}

The values for a just noticeable difference are then used to discretize the augmented parameter grids corresponding to images $1, 2$ and $3$. See table \ref{table:grids} for a summary of the grid sizes. A subset is drawn from each grid following the idea described in chapter \ref{chapter:stimuli}. Scaling the parameters of set $1$ with the just noticeable difference of $0.984$ results in a subset of $15$ images out of the original $23$ (figure \ref{fig:leaf_psych_set}). Set $2$ remains as is, as a just noticeable difference could not be defined. The just noticeable difference of $0.956$ that was found for set $3$ results in a subset of only $8$ images out of the original $22$ (figure \ref{fig:moss_psych_set}). These sets are optimized for visual testing in the sense that any consecutive images are just sufficiently similar. All sets are openly available on Zenodo \cite{zenodo}. 

\bigskip

\section{Discussion and conclusion} \label{chapter:conclusion}

Based on the results of this study, we can conclude that parameter grids based on perceived image quality can be constructed by applying psychometric scaling. The grids and threshold values thereby computed are unique to the base image used, and a value connecting computational similarity indices to a just noticeable difference cannot be determined in the general case. However, for building a dataset from one base image, the method described in this article is quite simple to execute. Analyzing the psychometric curves is rather easy and straightforward, as the numerical evaluations can be read directly from the graph. This scaling method --- at least in its current stage --- is best suited for use as a tool for dataset optimization, and to support or offer an alternative to rigorous mathematical research. The limitations of any attempt at psychological modeling must be recognized when applying to other fields of scientific research.

Given the novelty of the experiment and the moderate--sized survey data, some inherent uncertainty factors should be addressed. Additionally, given that the subject of this study deals in psychology as well as mathematics, a degree of uncertainty is to be assumed. The individual and fluctuating nature of perception and the high probability of mistakes in given evaluations add to this. Thurstone's paper \cite{thurstone} provides formulas for computing errors and deviations for given scales. However, they are not directly applicable for the study at hand, and their use should be postponed for further comparative studies in this subject. In his book \cite{engeldrum}, Engeldrum suggests the use of an uncertainty interval in analysis, to which the survey data that fall between the $0.25$ and $0.75$ marks belong to. The researcher could then state that when more than $75$ per cent of the observers provide a yes--answer, the desired quality is detected with certainty, and when less than $25$ per cent of the observers do so, the quality is certainly not detected. However, as we can see in figures \ref{fig:1d_leaf}, \ref{fig:1d_down} and \ref{fig:1d_moss}, each psychometric curve we have constructed falls entirely under the value of $0.75$. This is due to the accentuated variance in answers due to the sample size; as more data is gathered, the error grows smaller. In a pilot study we allow more flexibility in assessing the results as we would in a larger--scale study. Since we are using psychometric scaling for the construction of datasets to be used in further studies, there is no strong purpose for a margin of error to be this large. Some sanity checks were conducted in the analysis phase, and different groupings and methods of data fitting were tested. The results were all consistent with one another.

Given the resources to gather a large amount of data over the whole parameter grid at once, we would expect the results to be similar to those of this pilot study, but with a smaller margin of error. It should also be noted that the choice of the degree of the polynomial may have an effect on the results. The choice was made here to use third degree polynomial curves as they provided the best fit visually besides following the lead of Engeldrum \cite{engeldrum}.


The primary results of this study are the three psychometrically scaled, openly available datasets that are ready to use and optimized for visual testing in future studies. Secondly, a baseline for how perception--based testing in mathematical research can be applied is established in terms of methods and procedure. Furthermore, a lot of new information was learned about the link between the mathematics of imaging and the perception of visual data.

An obvious drawback of this scaling method is that the scaling and comparison testing must be done individually to each image used. However, once the set up for the experiment exists, it is fairly fast and easy to compute. The most prominent advantage of the method is related to the drawback; it is highly adaptable and can be easily personalized to fit an imaging task. A number of other ``-nesses'' than noisiness - such as focus or color intensity - could be surveyed, analysis could be computed using one reference image, another denoising method than TV could be used, or HaarPSI could be replaced by SSIM or any other quantifiable image property.

The experiments described here can be expanded for further testing with a wider audience to gather more data, but the psychometric scales as they are can be used as is to aid in subsequent studies in image quality and perception. The motivation behind this experiment campaign was to construct materials for research into preferred parameters in image denoising, which is the next objective of the research team. Sets $1$ and, especially, $3$ are strongly suited for said purpose. The availability of multiple sets for such a study allows for researching whether preferred parameters are also unique to a measured image or not. Additionally, the research methods developed during the course of this study have paved way for all ensuing research projects using the same data.

\section{Funding}

This research was supported by the Finnish Ministry of Education and Culture’s Pilot for Doctoral Programmes (Pilot project Mathematics of Sensing, Imaging and Modelling). The work of E.B. was supported by the Research Council of Finland through the Flagship of Advanced Mathematics for Sensing, Imaging and Modelling (decision number 359183). The work of M.J. and S.S. was supported by the Research Council of Finland through the Centre of Excellence of inverse modelling and imaging (decision number 353097) and the Flagship of Advanced Mathematics for Sensing, Imaging and Modelling (decision number 359182).

\begin{figure}[ht]
\centering
\begin{subfigure}{.5\textwidth}
  \centering
  \includegraphics[width=.8\linewidth]{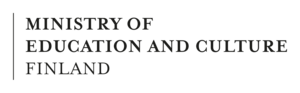}
\end{subfigure}%
\begin{subfigure}{.5\textwidth}
  \centering
  \includegraphics[width=.8\linewidth]{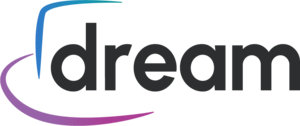}
\end{subfigure}
\end{figure}

\printbibliography

\end{document}